\documentclass[12pt]{JHEP3}
\input epsf.tex
\epsfclipon

\usepackage{epsfig}

\usepackage{bbm,bm,amsmath,amssymb}

\def\be{\begin{equation}}
\def\ee{\end{equation}}
\def\baray{\begin{eqnarray}}
\def\earay{\end{eqnarray}}

\newcommand{\roughly}[1]{\mathrel{\raise.3ex\hbox{$#1$\kern-0.85em
\lower1ex\hbox{$\sim$}}}}
\def\lsim{\roughly<}

\def\2pi{\left(2\pi\right)}

\def\sgn{{\rm sgn}}
\def\beq{\begin{equation}}
\def\eeq{\end{equation}}
\def\beqa{\begin{eqnarray}}
\def\eeqa{\end{eqnarray}}

\title{Overproduction of cosmic superstrings}

\author{Neil Barnaby, Aaron Berndsen, James M.\ Cline, Horace Stoica\\
Physics Department, McGill University \\
3600 University Street, Montr{\'e}al \\
        Qu{\'e}bec, Canada, H3A 2T8}

\date{\today}

\abstract{We show that the naive application of the Kibble mechanism seriously
underestimates the initial density of cosmic superstrings that can be formed during the
annihilation of D-branes in the early universe, as in models of brane-antibrane	
inflation.   We study the formation of defects in effective field theories of the string
theory tachyon  both analytically, by solving the equation of motion of the tachyon
field near the core of the defect, and numerically, by evolving the tachyon field on a
lattice. We find that defects generically form with correlation lengths of order
$M_s^{-1}$ rather than $H^{-1}$.  Hence, defects localized in extra dimensions may
be formed at the end of inflation.  This implies that brane-antibrane inflation models
where inflation is driven by branes which wrap the compact manifold may have problems
with overclosure by cosmological relics, such as domain walls and monopoles.}

\maketitle

\begin{document}

\section{Introduction}
\label{intro}

Although it is notoriously difficult to test string theory in the laboratory,
exciting progress has been made on the cosmological front through the
realization that superstrings could appear as cosmic strings within recent
popular scenarios for brane-antibrane inflation
\cite{Tye:StringProduction}-\cite{CosmicStrings}.  Gravity wave detectors and 
pulsar  timing measurements could thus give the first positive experimental
signals coming from  superstrings \cite{GW} (see, however, \cite{SS}). If seen,
it will be challenging to distinguish cosmic superstrings from conventional
cosmic strings.  One distinguishing feature is the possibility that
superstrings have a smaller intercommutation probability than ordinary cosmic
strings \cite{JJP}.  In this paper we consider whether the mechanism of
formation of cosmic superstrings might provide another source of distinction,
by studying in detail the formation of string and brane defects, 
emphasizing differences with the standard picture of defect formation.

To set the stage, let us briefly recall how  defect formation works in a standard
theory \cite{Kibble} (see, for example \cite{Brandenberger}, for modern review).  
Consider a scalar field theory with potential $\frac{\lambda}{4}(|\phi|^2-\eta^2)^2$. 
In the standard  picture as the universe cools below some critical temperature $T_c$
the $U(1)$ symmetry is broken and $\phi$ rolls to the degenerate vacua of the theory
$|\phi|=\eta$.   Causality implies that the field cannot be correlated throughout the
space and hence the field rolls to different vacua in different spatial regions leading
to the formation of topological defects. The defect separation is set by the
correlation length, $\xi$.  For a universe expanding with Hubble rate $H$ one expects
$\xi < H^{-1}$ and hence a minimum defect density of about one per Hubble  volume. 
However, this is only an upper bound on $\xi$.  A more careful estimate can be made
using  condensed matter physics methods: equating the free energy gained by symmetry
breaking with the gradient  energy lost.  Very close to $T_c$ thermal fluctuations
which can restore the symmetry are probable; however, once the universe cools below the
Ginsburg temperature, $T_G$, there is insufficient thermal energy to  excite a
correlation volume into the state $\phi=0$ and the defects ``freeze out.''\ \  For the
scalar field theory described above this estimate yields a correlation length of order
the microscopic scale:  $\xi \sim \lambda^{-1} \eta^{-1}$.

The physics of tachyon condensation on the unstable D$p$ brane-antibrane system is quite different however,
due to the peculiar form of the action \cite{Sen:TachyonMatter}
\beq
	S = -2 \tau_p \int d^{p+1} x\, e^{-|T|^2/\,b^2}\, \sqrt{1+|\partial T|^2}.
\eeq
The tachyon potential has a ``runaway'' form and there are no oscillations of the field near the true vacuum
which can restore the symmetry.  The decaying exponential potential of the complex tachyon field multiplies 
the kinetic term.  Once condensation starts, $T$ rolls quickly to large value, and damps the gradient energy
exponentially.  This essentially eliminates the restoring force which would tend to erase gradients within a
causal volume in an ordinary field theory.  

In this paper we perform a quantitative analysis of the formation of string
defects starting from the unstable tachyonic condensate that describes unstable
brane-antibrane systems.  Having established that defects form with an energy 
density comparable to that which is available from the condensate, we then examine
the possible cosmological consequences of this larger-than-expected initial
density.

The organization of this paper is as follows.  
In section \ref{BraneInflation} we
briefly discuss  brane-antibrane inflation and the formation of defects at the
end of brane-antibrane inflation by tachyon condensation.
In section \ref{kinkSection} we study the formation of codimension-one branes in the decay of
a nonBPS brane both analytically as well as through lattice simulations. In
section \ref{vortexSection}  we repeat the study, this time for the formation of
codimension-two branes in the decay of the brane-antibrane system. 
We contrast these results with the formation of conventional cosmic strings in 
section \ref{sect5}.  
The reader who is interested in the cosmological implications may wish to skip
directly to section \ref{Consequences} where we consider new constraints on brane
inflation models which may arise due to the large initial density of defects.
Conclusions are given in section \ref{Conclusions}.  In an appendix we justify the
assumed initial conditions for the tachyon fluctuations which lead to defects.

\section{Brane Inflation}
\label{BraneInflation}

Relic cosmic superstrings can form at the end of inflation driven by the 
attractive interaction of branes and antibranes \cite{BraneInflation}. In this
picture, inflation ends  when the brane and antibrane (or a pair of branes
oriented at angles) become sufficiently close that one of the open string modes
stretching between the branes becomes tachyonic.  The subsequent rolling of
this tachyon  field describes the decay of the brane-antibrane pair. Quantum
fluctuations produce small inhomogeneities in the tachyon field, which will
cause it to roll toward different vacua in  different spatial regions,  leading
to the formation of topological defects which are known to be consistent
descriptions of lower dimensional branes.

The branes which drive inflation must span the three noncompact dimensions and may wrap
some of the compact dimensions.  The defects which form are lower dimensional branes
whose world-volume is within the world-volume of the parent branes.  The argument was
made in \cite{Tye:StringProduction} (illustrated in figure \ref{kibble_arg}) that by 
applying the reasoning of the  Kibble mechanism to the formation of the
lower-dimensional branes one  concludes that the branes which are  copiously produced
as topological defects will always wrap the same  compact dimensions that the parent
branes wrap and hence these branes appear as cosmic strings to the  $3$-dimensional
observer.\footnote{The production of phenomenologically dangerous monopole-like and
domain  wall-like defects is suppressed.}\   This argument is based on the fact that
the size of the compact dimensions is orders of magnitude smaller than the Hubble
distance during (and at the end of) inflation and hence there are no causally
disconnected regions along the compact dimensions (a reasonable estimate is $H^{-1}
\sim 6 \times 10^{3} M_s^{-1}$ while typically the extra dimensions are compactified at
a scale $R$ closer to $M_s^{-1}$).  Therefore along these directions   the tachyon
field would always roll toward the same vacuum  and no topological defects would form.
The causally  disconnected regions occur only along the extended  directions, so the
defects are localized along the extended directions.  An identical argument implies
that cosmic strings should form with a density of about one string per Hubble  volume. 

\EPSFIGURE[ht]{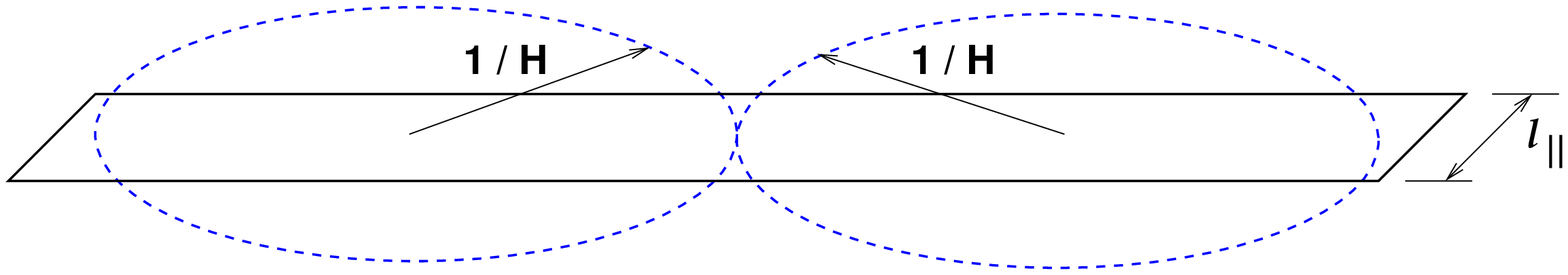,width=5in}{Illustration of the Kibble argument in the case of branes spanning 
extended
dimensions and wrapping compact ones. The compact dimension is much smaller in size than the
Hubble distance,
$1/H \gg \ell_{\parallel}$, so there are no causally disconnected regions along this dimension. The
branes that will form as topological defects will wrap the same compact dimensions as the parent 
branes, and will be localized on the extended dimensions.\label{kibble_arg}}

Reference \cite{Tye:StringProduction} uses the analogue of the condensed matter physics
argument paraphrased in section \ref{intro} to estimate the correlation length of the
tachyon field at the end of inflation.   Since the universe has zero temperature at the
end of inflation the thermal fluctuations are replaced by the quantum fluctuations of
de Sitter space: $H/(2 \pi)$.  In that analysis the potential depends on the brane
separation and as a result of the relative motion of the brane and the antibrane,  the
curvature of the potential at $T=0$ changes sign.  It is also assumed that the tachyon
potential, $V(T)$, has a minimum for some finite value of $T$.  In this case there is
an interplay between the  correlation length, $\xi_C$ given by the curvature of the
potential at  $T=0$, and the Ginsburg length,  $\xi_G$, which is the largest scale on
which quantum fluctuations of the field can  restore the symmetry. As the brane
separation decreases and the shape of the potential changes,  the correlation length
and the Ginsburg length change as well. The density of defects freezes when  $\xi_C =
\xi_G$.  This analysis yields a correlation length which is not much different from the
Hubble scale: $\xi \sim 1.6 \times 10^{4} M_s^{-1} \sim 2.7 H^{-1}$.




There are several reasons to consider a more quantitative study of
defect formation at the end of brane inflation.  Firstly, the estimate of one
defect per Hubble volume provides only a lower bound on the defect density via
causality. The actual network of defects produced is determined by the
complicated tachyon dynamics as it rolls from the unstable maximum of its
potential to the degenerate vacua of the theory.  Second,
the effective field theory which
describes the dynamics of the tachyon \cite{Sen:TachyonMatter} has a rather
unusual causal structure  \cite{CarollianContraction} and the usual 
reasoning of the Kibble mechanism may not be applicable.  In  \cite{CarollianContraction} it
was found that in the case of the homogeneous rolling tachyon the small fluctuations of
the field propagate according to a ``tachyon effective metric'' which depends on
the rolling tachyon background.   As the tachyon rolls towards the vacuum the
tachyon effective metric degenerates, the tachyon light cone collapses onto a
time-like half line and the tachyon fields at different spatial points are
causally decoupled.

Finally we comment on the validity of the effective field theory description.   At
some  point the effective field theory description of the decaying brane will become
inappropriate and the correct degrees of freedom will be the decay products of the
brane annihilation.  Since the topological defects of the tachyon field form in a short
time of order $M_s^{-1}$,  \cite{Sen:TimeEvolution}-\cite{Reheating2}, the effective
field theory description should be applicable during the period of defect formation. 
Furthermore, there are reasons to believe that the effective field theory is valid
even at  late times when the tachyon is close to the vacuum, since in this regime the
tachyon effective field theory  may give a dual description of the closed strings which
are produced during brane decay \cite{ClosedStrings}.

\section{Tachyon Kink Formation in Compact Spaces}
\label{kinkSection}

In this section we consider the formation of codimension-one branes from tachyon condensation on a nonBPS
brane.  We study defect formation both by solving the full nonlinear equations of motion on a 
lattice, as well as providing analytical approximations to the behavior of the tachyon field in different
regimes of interest.  We study the evolution of the tachyon field starting from a profile which is close to 
the unstable vacuum $T=0$ (in this respect our analysis is similar to 
\cite{TachyonicPreheating1,TachyonicPreheating2}).  In our analysis there is no parameter which causes 
continuous variation of the potential, we consider the brane-antibrane system to be coincident at $t=0$ when
the initial conditions are imposed (this potentially important difference should be kept in mind when 
comparing our results to \cite{Tye:StringProduction}).

\subsection{Action and Equation of Motion}

We will work with the tachyon Dirac-Born-Infeld effective action 
\cite{Sen:TachyonMatter,DBIAction,Sen:KinkAndVortex}

\begin{equation}
\label{kinkaction}
   S = -\int V(T) \sqrt{ 1 + \partial_\mu T \partial^\mu T} \, \sqrt{-g} \, d^{\,3+1+1}x.
\end{equation}
The action (\ref{kinkaction}), with $V(T) = V_0 \, / \cosh(T/\sqrt{2 \alpha'})$, can be derived from string 
theory in some limit \cite{DBIderive} and has been shown to reproduce 
various nontrivial aspects of the full string theory dynamics \cite{Sen:KinkAndVortex,DBIproperties}.  Here
we take the potential to be 

\begin{equation}
\label{potential}
V(T) = \tau_{p} \, \exp\left(-T^2 / \, b^2\right)
\end{equation}
where $\tau_p$ is the D$p$-brane tension. The constant $b$ determines the tachyon mass in the perturbative 
vacuum as $M_T = \sqrt{2} b^{-1} \sim 1/\sqrt{\alpha'}$.  Qualitatively the results will depend very little 
on the specific functional form of $V(T)$.  In fact, for much of the analytical analysis which follows we 
will not even need to make reference to the specific functional form of 
$V(T)$.\footnote{Provided of course that $V'(T=0)=0$ and $V(T \rightarrow \pm \infty)
\rightarrow 0$.}\ \   
We consider real tachyon fields in a spacetime with three expanding, noncompact dimensions $\{x^i\}$ (where
$i = 2,3,4$) and one compact dimension $x^1 = x$ which we take to be static.  The metric is
\begin{equation}
\label{kinkMetric}
  g_{\mu\nu}dx^{\mu}dx^{\nu} = -dt^2 + dx^2 + a(t)^2\delta_{ij}\,dx^{i}\,dx^{j}.
\end{equation}
For simplicity we take the tachyon field to depend only on time, $x^0=t$, and the compact spatial coordinate
$x^1 = x$.  The equation of motion which follows from (\ref{kinkaction}) with the metric (\ref{kinkMetric}) 
is

\begin{equation}
\label{dampedKinkEOM}
  \left( 1 + T'^2 \right) \ddot{T} = T'' \left( 1 - \dot{T}^2 \right) - \left(3 H \dot{T} 
  + \frac{V'(T)}{V(T)} \right) \left( 1 - \dot{T}^2 + T'^2 \right) + 2\,\dot{T}\, T'\, \dot{T}'
\end{equation}
where $\dot{T} = \partial_0 T = \partial_t T$, $T' = \partial_1 T = \partial_x T$, 
$V'(T) = \partial V / \partial T$ and $H = \dot{a} / a$.

\subsection{Lattice Simulation of Kink Formation}
\label{kinkNumerical}

We have solved (\ref{dampedKinkEOM}) on a lattice for different values of the compactification radius and
the Hubble parameter $H$ (which we take to be constant for simplicity).  We consider vanishing initial 
velocity
$\dot{T}(t=0,x) = \dot{T}_i(x) = 0$.  The initial profile $T(t=0,x) = T_i(x)$ is a Fourier series truncated 
such that the minimum wavelength is comparable to the lattice spacing.\footnote{See the appendix for a 
discussion of the validity of these initial conditions.}  The amplitudes of the Fourier 
coefficients are given by a random Gaussian distribution and the overall amplitude of the initial profile is
chosen to be small compared to $b$.

As in \cite{Cline:DbraneCondensation} we find that the gradient of the tachyon field near the core of the
kink becomes infinite in a finite time, forcing us to halt our evolution.  After this point the
codimension-one branes have formed and if one wants to follow the evolution beyond this time it is necessary
to consider the dynamics between branes and antibranes in a compact space; the field theory description is
no longer adequate.

Typically the initial profile crosses $T=0$ at many points.  In the early stages of the evolution the
field begins to grow due to the small displacement from the unstable vacuum.  During this phase of the
evolution many of the small fluctuations of the field will straighten themselves out.  Large Hubble
damping tends to kill off the high frequency fluctuations faster.  At the end of this initial stage there
may be several locations where the field stays pinned at $T=0$, depending most crucially on the radius of
compactification.  The evolution very quickly enters a nonlinear regime in which the field begins to roll
quickly towards $T \rightarrow +\infty$ where $T>0$ and $T \rightarrow -\infty$ where $T<0$.

The most important parameter for determining the density of the defect network once the 
daughter branes have formed
is the radius of compactification, $R$.  Perhaps surprisingly, we find that for a compactification radius as 
small as a few times $b$, tachyon kinks will form.  Once the field enters the nonlinear regime the defect 
formation depends only on the local physics near the core of the kink.  In figures \ref{kinkFig1},
\ref{kinkFig2} we plot the tachyon field versus $t$ and $x$, showing the 
formation of codimension-one branes from the decay of a nonBPS brane for radii of compactification
$R = 8 M_T^{-1}$ and $R = 15 M_T^{-1}$.  The Hubble constant is taken to be vanishing, $H=0$,
in these figures.  We have also considered $H \not= 0$ and find that the Hubble damping has little effect on
the final kink/anti-kink network.  Hence we find that tachyon kinks \emph{do} form in the compact directions
even when the field is initially in causal contact throughout the extra dimension. 

\DOUBLEFIGURE[ht]{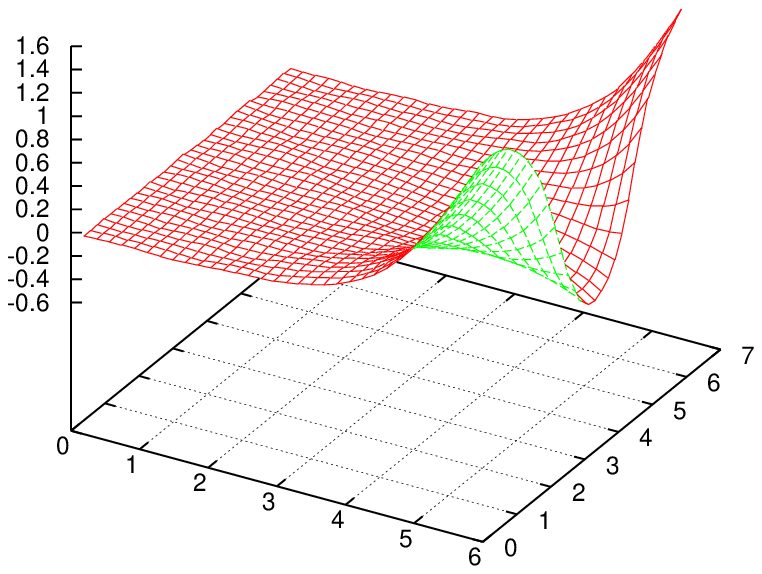,width=2.75in,height=2in}{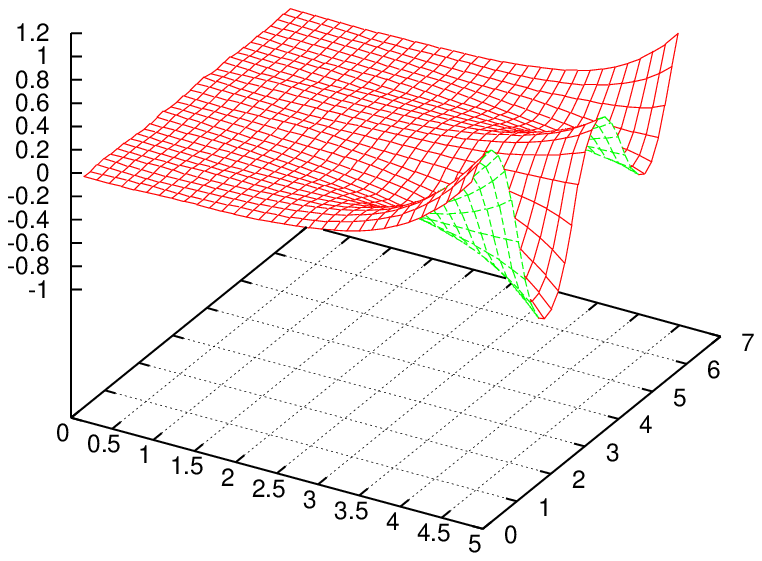,width=2.75in,height=2in}{Formation of tachyon kinks for $R = 8 M_T^{-1}$.  The time axis is measured in units of $M_T^{-1}$. \label{kinkFig1}}{Formation of tachyon kinks for $R = 15 M_T^{-1}$.  The time axis is measured in units of $M_T^{-1}$. \label{kinkFig2}}

\subsection{Analytical Study of Tachyon Kink Formation}
\label{kinkAnalyticalSection}

Here we describe analytically the formation of tachyon kinks.  Since the full equation of motion 
(\ref{dampedKinkEOM}) is difficult to solve analytically for arbitrary initial data, we study the dynamics
of (\ref{dampedKinkEOM}) in different regimes: near the core of the defect, where the field stays pinned at
$T=0$, and away from the core, where $T$ rolls towards the vacuum.  To simplify our analysis we neglect
the compact topology of the extra dimension $x$, which should be a good approximation since the defect solutions
are infinitely thin and therefore highly localized.

\subsubsection{Solutions Near the Core of the Defect}
\label{kinkCoreSolutions}

Here we briefly review the analytical studies of kink formation near the core of the defect presented
in \cite{Reheating2}.  Consider initial data $T(t=0,x)=T_i(x)$ and 
$\dot{T}(t=0,x)=\dot{T}_i(x)=0$.  The field should start to roll where $T_i(x) \not= 0$ due to the 
small displacement from the unstable maximum $V'(T)=0$, and furthermore it must stay pinned 
at $T=0$ at the core of the kink.  At $t=0$ the equation of motion 
(\ref{dampedKinkEOM}) is

\[
\ddot{T}_i(x) \left(1+T'_i(x)^2\right) = T''_i(x) + \frac{2}{b^2}T_i(x)\left( 1 + T'_i(x)^2
\right).  
\]
Clearly any point $x_0$ where $T_i(x_0)=T_i''(x_0)=0$ will be a fixed point where 
$\ddot{T}(t,x_0)=0=\dot{T}(t,x_0)$ throughout the evolution.  We restrict ourselves to 
considering only 
initial data such that $\sgn(\ddot{T}_i(x)) = \sgn(T_i(x))$ for all $x$ to ensure that the solutions are 
increasing.  To analytically study the dynamics near $x_0$ it is therefore
reasonable to make the ansatz
\begin{equation}
\label{kinkAnsatz}
  T(t,x) \cong u(t) (x-x_0).
\end{equation}
For $u(t) = \mathrm{const} \rightarrow \infty$ this solution corresponds to the singular static soliton
solution of Sen \cite{Sen:KinkAndVortex}.  Inserting the ansatz (\ref{kinkAnsatz}) into (\ref{dampedKinkEOM})
and working only to linear order in $(x-x_0)$ one obtains an ordinary differential equation for the slope

\begin{equation}
\label{smalldamped1}
  \ddot{u} = \frac{2}{b^2} u + 2 \frac{u \dot{u}^2}{1+u^2} - 3 H \dot{u}.
\end{equation}
We have solved (\ref{smalldamped1}) numerically for both constant $H$ and $H \sim 1/(t+t_0)$.  We find that
generically the solutions of (\ref{smalldamped1}) become singular in a finite time $t=t_c$.  Larger $H$ tends
to delay the onset of the singularity and soften the singular behaviour.  To understand this finite time
slope divergence analytically it is simplest to consider $H=0$.  In this case if we assume that initially 
$\dot{u}$ is close to zero then the second term on the right hand side of (\ref{smalldamped1}) can be 
neglected and therefore

\[
  u(t) \cong u_{+} e^{\sqrt{2}t/b} + u_{-} e^{-\sqrt{2}t/b}
\]
at early times.  Clearly $\dot{u}$ grows quickly and the second term on the right hand side of 
(\ref{smalldamped1}) very quickly becomes important.  In the regime where $u$ and its derivatives become 
large the solution has the behaviour

\[
  u(t) \sim \frac{k}{t_c-t}
\]
where the critical time $t_c$ depends on the initial conditions.  
We find that within finite time the slope
becomes singular and the time dependent tachyon field near the core of the kink coincides with the singular
soliton solution of Sen \cite{Sen:KinkAndVortex}.  Hence we conclude that the codimension-one brane forms 
in a finite time.

\EPSFIGURE[ht]{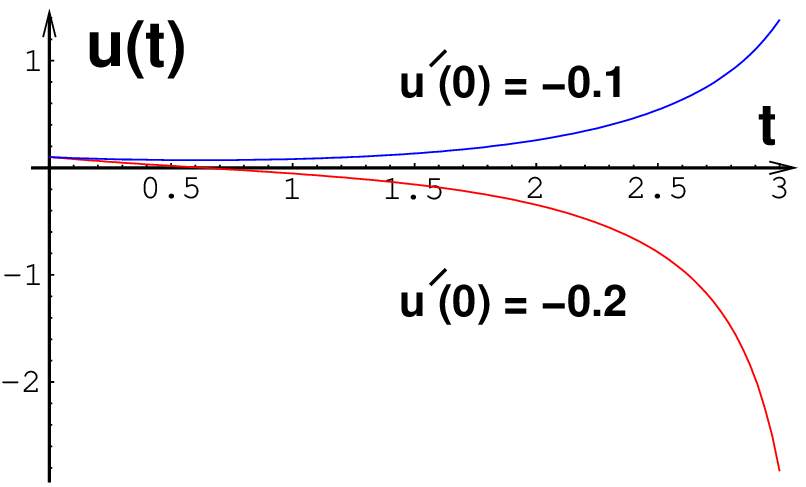,width=3.75in,height=1.75in}{Behavior of the slope at the core of the 
kink for different initial conditions.  The time axis is measured in units of $b$.\label{kinkSlopeFig}}

This finite-time slope divergence was observed both numerically and analytically
in \cite{Cline:DbraneCondensation} and leads to a finite-time divergence in the stress-energy tensor.
This divergence was also found in an exact string theory calculation in \cite{Sen:TimeEvolution}.
For the potential (\ref{potential}) the stress-energy tensor near $x=x_0$ and $t=t_c$, with $H=0$, is 
\cite{Reheating2}

\[
 T_{00} \cong \frac{\tau_p k}{t_c-t}\exp\left( -\frac{k^2 (x-x_0)^2}{b^2(t_c-t)^2} \right).
\]
Hence, with the normalization $b = 2 \sqrt{\pi \alpha'}$ proposed in \cite{Sen:KinkAndVortex}, as 
$t \rightarrow t_c$ we have $T_{00} \rightarrow \tau_{p-1} \delta(x-x_0)$.  Similarly one can
show that in this regime $T_{01} = T_{11} = 0$ and that $T_{ij} = -\tau_{p-1} \delta_{ij} \delta(x-x_0)$.  
In the limit of condensation, then, the stress-energy near $x=x_0$ is identical to that of a 
D$(p-1)$-brane.  If we take into account the rolling of the tachyon for $x \not= x_0$ then there will be an 
additional component to $T_{\mu\nu}$ corresponding to tachyon matter, as in 
\cite{Ishida:RollingDownToDbrane}.

\subsubsection{Solutions Near the True Vacuum}

Away from the site of the kink the field will roll towards the true vacuum 
$T \rightarrow \pm \infty$ so that $V(T) \rightarrow 0$ at late times for $x \not= x_0$.
As $V(T) \rightarrow 0$ one can show both analytically \cite{VacuumDBI} and numerically 
\cite{Caustics1} that (\ref{dampedKinkEOM}) has an attractor which satisfies

\begin{equation}
\label{eikonal}
  \partial^\mu T \partial_\mu T + 1 = 0.
\end{equation}
This attractor holds for arbitrary metric and hence the solutions we describe below hold equally well
for vanishing or nonvanishing $H$.  Exact solutions of (\ref{eikonal}) were found 
in \cite{Caustics1} for arbitrary initial data.  The solution is given along a set of characteristic curves
which, in $1+1$ dimensions, are defined by

\begin{equation}
\label{characteristiccurve}
  x(q,t) = q - \frac{T_i'(q)}{\sqrt{1+T_i'(q)^2}}\,t
\end{equation}
where the parameter $q$ defines the initial position of the curve on the $x$ axis such that $x(q,t=0)=q$ and 
$T_i(x)$ is the Cauchy data at time $t=0$.  The parameter $q$ should be thought of as labeling the curves.  
The value of the tachyon field along a given characteristic curve is

\begin{equation}
\label{characteristicT}
  T(q,t) = T_i(q) + \frac{t}{\sqrt{1+T_i'(q)^2}}.
\end{equation}
For generic initial data these solutions form caustics in finite time where the second derivatives of the 
field blow up and the effective description (\ref{kinkaction}) breaks down.  In the present context the 
caustics lead to singularities in the stress-energy tensor \cite{Caustics2} for $x \not= x_0$.  Caustic 
formation does not arise in our lattice simulations since the finite-time first derivative divergence 
associated with defect formation forces us to halt our evolution, and this generically seems to 
occur before caustic formation.  The formation of caustics is predicted by a wide range of tachyon effective 
field theories \cite{Caustics3} and is presumably not present in the full string theory dynamics.

\section{Tachyon Vortex Formation in Compact Spaces}
\label{vortexSection}

In this section we generalize the results of section \ref{kinkSection} to consider the formation of 
codimension-two branes from tachyon condensation on the brane-antibrane pair.  This situation is of more
direct interest to brane-antibrane inflation since inflation ends when a brane-antibrane pair annihilate.
This situation is also more realistic since the stable branes in a given string theory are those whose 
dimension differs by a multiple of two.

\subsection{Action and Equations of Motion}
\label{vortexActionSection}

We wish to generalize the results of the previous section to consider complex tachyon fields which depend
on time and two spatial coordinates which we compactify on a square torus.  We generalize the action 
(\ref{kinkaction}) to

\begin{equation}
\label{vortexaction}
S = -2 \tau_p \int  e^{-|T|^2 / \, b^2} \, \sqrt{ 1 + \partial_\mu T \partial^\mu \bar{T}} 
     \, \sqrt{-g} \, d^{3+2+1}x
\end{equation}
with metric
\[
  g_{\mu\nu} dx^{\mu}dx^{\nu} = -dt^2 + dx^{2} + dy^{2} + a(t)^2\delta_{ij} dx^i dx^j
\]
where $x^1 = x$ and $x^2 = y$ are Cartesian coordinates on the torus and $x^i$ denotes the three noncompact
dimensions, as in section \ref{kinkSection}.  We consider tachyon fields which depend on $x^0=t$, $x^1=x$ and
$x^3=y$.  If we separate the tachyon field into into real and imaginary components as
\[
  T(t,x,y) = T_1(t,x,y) + i \, T_2(t,x,y)
\]
then (\ref{vortexaction}) may be rewritten in terms of components as
\[
  S = -2 \tau_p \int \exp \left( -T_J T_J / \, b^2 \right) \sqrt{ 1 + \partial_\mu T_I \partial^\mu T_I} 
  \, \sqrt{-g} \, d^{3+2+1}x
\]
where the upper-case Roman indices label the real and imaginary components of the tachyon field ($I=1,2$).
The equation of motion is (no sum over $I$, sum over $J$ and $K$):
\begin{eqnarray}
&& V\left(|T|\right)\partial_{\alpha}
\left[\frac{\sqrt{-g}g^{\alpha\beta}\partial_{\beta}T_I}
{\sqrt{1+\partial_{\alpha}T_J\partial^{\alpha}T_J}}\right]
+\frac{\partial V\left(|T|\right)}{\partial T_K}
\left[\frac{\sqrt{-g}\partial_{\alpha}T_K\partial^{\alpha}T_I}
{\sqrt{1+\partial_{\alpha}T_J\partial^{\alpha}T_J}}\right] \nonumber  \\
&& -\frac{\partial V\left(|T|\right)}{\partial T_I}\sqrt{-g}
\sqrt{1+\partial_{\alpha}T_J\partial^{\alpha}T_J} = 0.
\label{vortexEOM}
\end{eqnarray}

We note that the action (\ref{vortexaction}) has not been derived from first principles and has several
drawbacks from a theoretical perspective.  For other proposals of the effective field theory on
the brane-antibrane pair see \cite{Sen:KinkAndVortex},\cite{Tye:ImprovedAction}-\cite{Garousi:VortexAction}.
One theoretical difficulty is that the action (\ref{vortexaction}) does not evade Derrick's theorem 
\cite{StaticKinks}.  This is of no practical concern to us since we study time dependent solutions and since
our interest is in defect formation in a compact space where charge conservation precludes the possibility 
of isolated defect solutions. More seriously, for the action (\ref{vortexaction}) a static profile of the 
form $T = u(x + i y)$ with $u \rightarrow \infty$ does not correctly reproduce the stress tensor for a 
codimension-two brane.  We will attempt to address this difficulty by considering
an alternative description of the complex tachyon dynamics in a subsequent subsection.

\subsection{Lattice Simulations of Vortex Formation}
\label{vortexNumericalSection}

We solve the system of two coupled partial differential equations (\ref{vortexEOM}).  As in subsection 
\ref{kinkNumerical} we find that the gradient of tachyon field becomes singular near the core of the defect 
in a finite time, forcing us to halt our lattice evolution.  As in the case of the kink we choose as initial
data $T(t=0,x,y)=\dot{T}_i(x,y) = 0$ and $T(t=0,x,y)=T_i(x,y)$ given by a truncated Fourier series with 
random Gaussian coefficients with the overall amplitude small compared to $b$.  We take $H=0$ for our 
examples, since the Hubble damping plays little qualitative role in the dynamics.  In figures 
\ref{vortexFig0}, \ref{vortexFig1}, \ref{vortexFig2} and \ref{vortexFig3} we plot 
$-\left|T\right| = -\sqrt{T_1^2 + T_2^2}$ against $\{x,y\}$ with 
the $T$ axis measured in units of $b$.  Figure \ref{vortexFig0} shows typical initial
conditions used for our numerical analysis.  In figures \ref{vortexFig1}, \ref{vortexFig2} and 
\ref{vortexFig3} we plot the final configurations close
to $t=t_c$ when the gradients become infinite for various radii of compactification. (Note that because we 
are plotting $-|T|$ rather than $+|T|$ the vortices appear as spikes in the final configuration.)  Again we 
find that vortices do form for radii of compactification as small as a few times $M_T^{-1}$, 
even though the field 
is initially in causal contact throughout the compact space.

\DOUBLEFIGURE[ht]{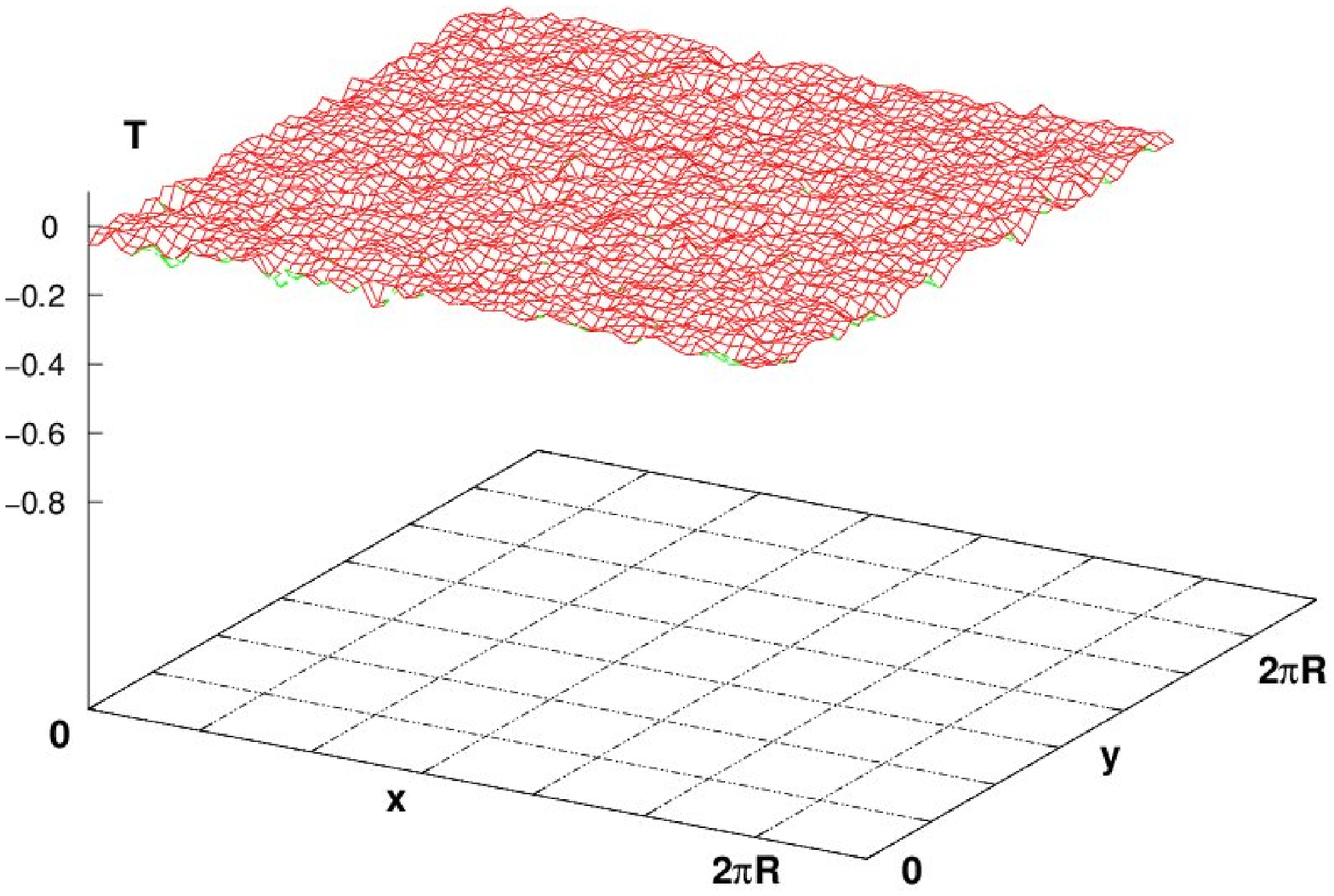,width=2.75in,height=2in}{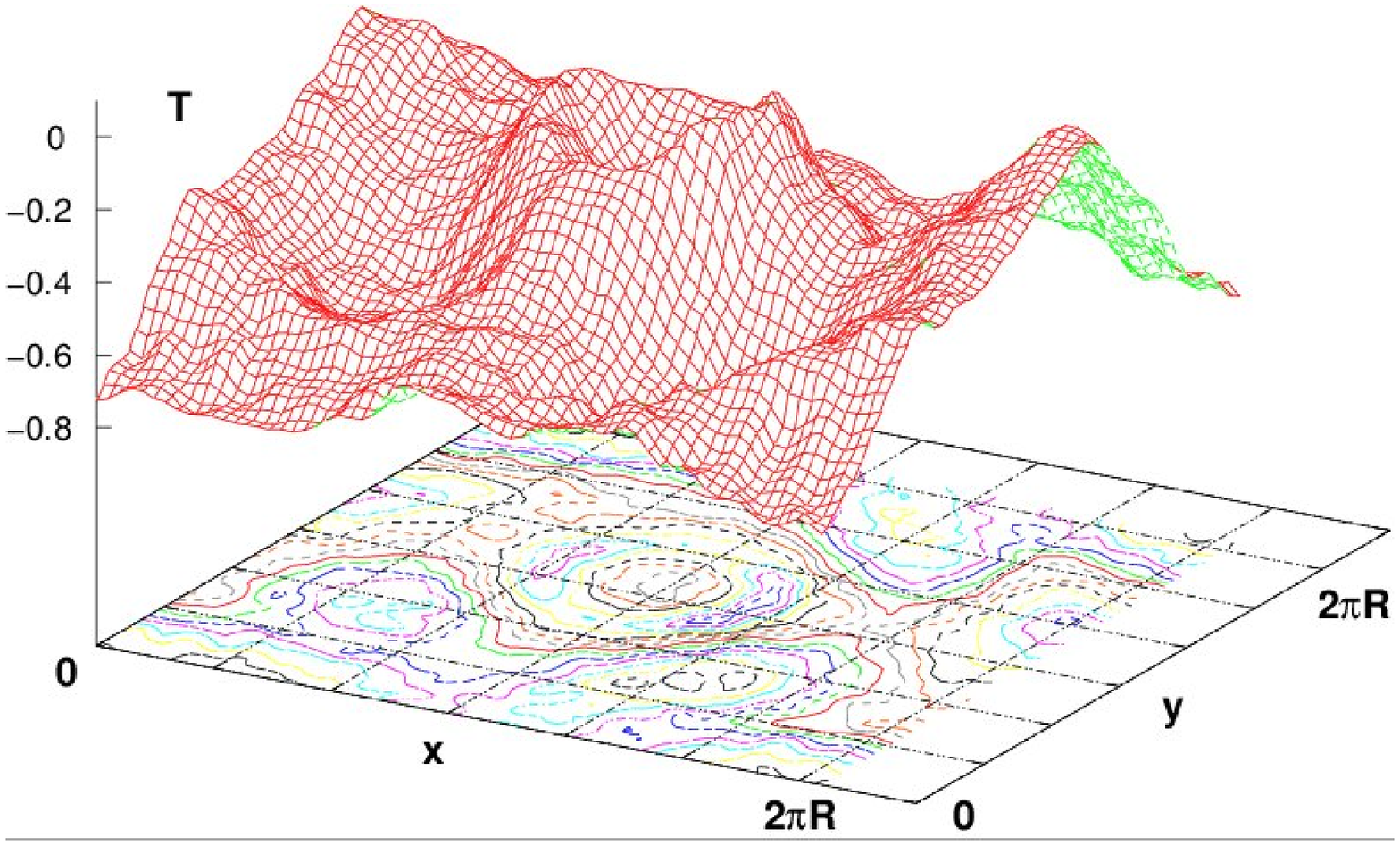,width=2.75in,height=2in}{Typical initial configuration for numerical studies of vortex formation. \label{vortexFig0}}{The final configuration of the complex tachyon field for $R = 7 M_T^{-1}$. \label{vortexFig1}}

\DOUBLEFIGURE[ht]{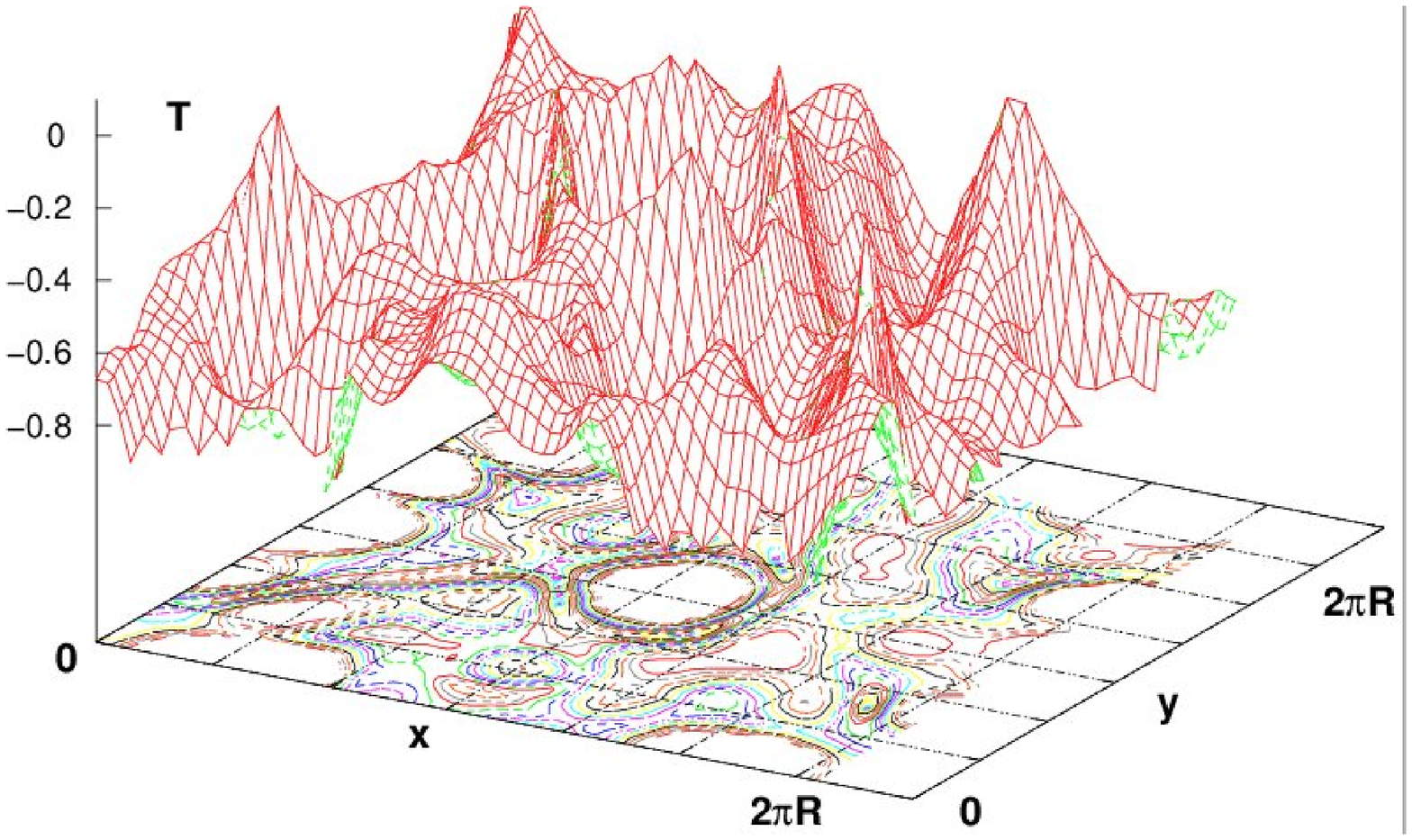,width=2.75in,height=2in}{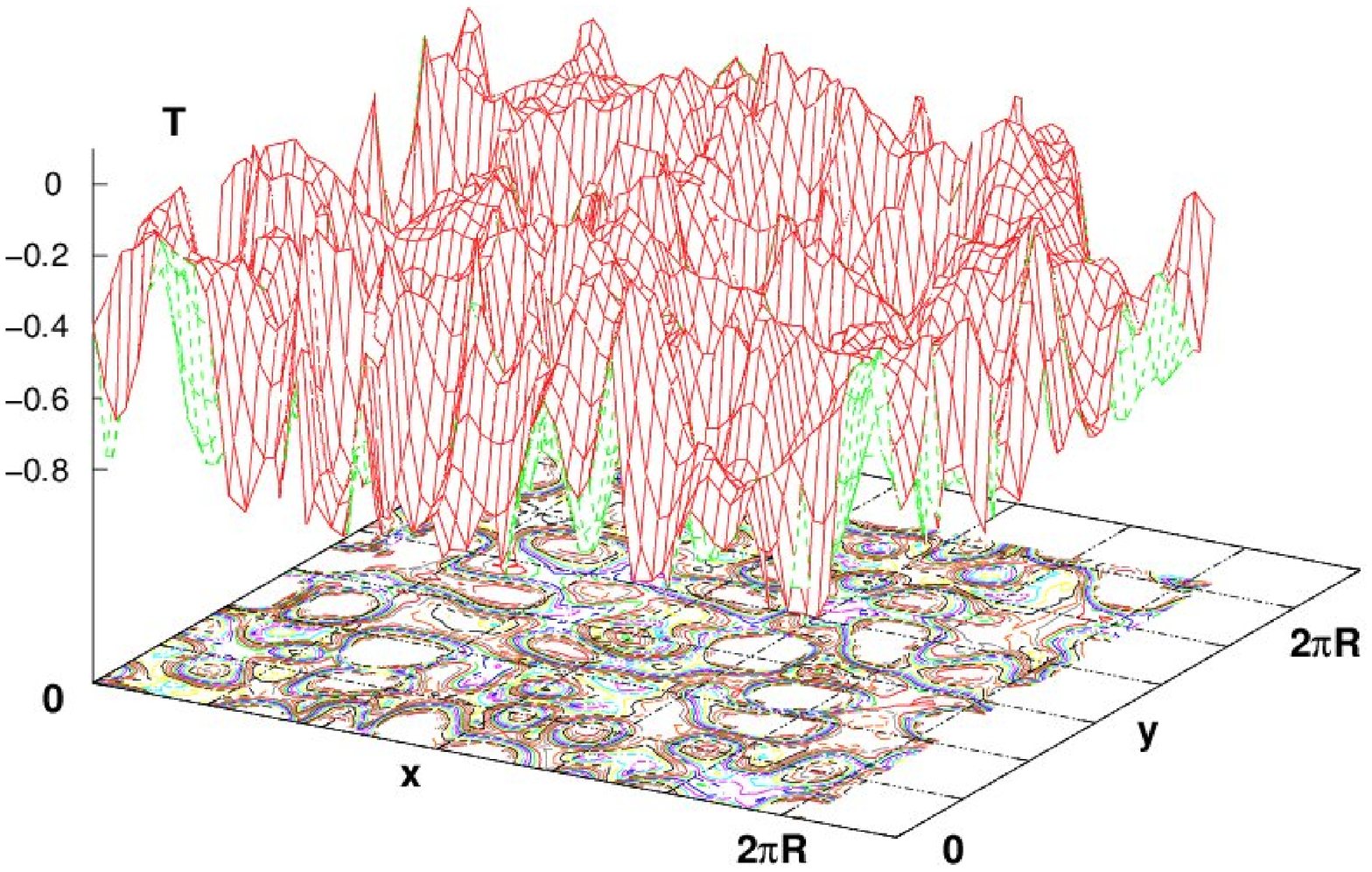,width=2.75in,height=2in}{The final configuration of the complex tachyon field for $R = 15 M_T^{-1}$. \label{vortexFig2}}{The final configuration of the complex tachyon field for $R = 25 M_T^{-1}$. \label{vortexFig3}}

\subsection{Analytical Study of Vortex Formation}

As in subsection \ref{kinkCoreSolutions} we expect that in the case of the vortex there will exist
fixed points where the field stays pinned at $T=0$ throughout the evolution.  To see this we consider the
equation of motion for $T_1$ at $t=0$ for initial data such that $\dot{T}(t=0,x,y) = 0$.  To simplify the
expression we write the equation evaluated at a point $(x_0,y_0)$ such that $T(t=0,x_0,y_0)=0$:
\begin{eqnarray*}
 &\left[ -\ddot{T}_1 + \partial_x^2 T_1 + \partial_y^2 T_1 \right]
 \left[\phantom{|_|^|\!\!} 1 + (\partial_x T_1)^2 + (\partial_y T_1)^2 + (\partial_x T_2)^2 + 
    (\partial_y T_2)^2 \right]& \\
 &-\left[\phantom{|_|^|\!\!} (\partial_x T_1)^2\, \partial_x^2 T_1 + (\partial_y T_1)^2\,
	 \partial_y^2 T_1 
  + 2\, \partial_x T_1\, \partial_y T_1\, \partial_x \partial_y T_1 \right.& \\
 & \left.\phantom{|_|^|\!\!} \partial_x T_1\, \partial_x T_2\, \partial_x^2 T_2 
 +\partial_y T_1\, \partial_y T_2 \,\partial_y^2 T_2
 + \left( \partial_x T_1\, \partial_y T_2 +  \partial_y T_1 \,\partial_x T_2 \right)
    \partial_x \partial_y T_2 \right] = 0.&
\end{eqnarray*}
The equation for $T_2$ is identical with $1 \leftrightarrow 2$.  It is clear, then, that any point 
$(x_0,y_0)$ where 
$T_I(0,x_0,y_0) = \partial_x^2 T_{I}|_{(0,x_0,y_0)} = \partial_y^2 T_{I}|_{(0,x_0,y_0)} = 
\partial_x \partial_y T_{I}|_{(0,x_0,y_0)} = 0$
(for $I=1,2$) will be a fixed point of the evolution where $\ddot{T}_I(t,x_0,y_0) = \dot{T}_I(t,x_0,y_0) = 0$
for all $t$ and hence the field $T$ stays pinned at zero throughout the evolution.  In the neighborhood
of the point $(x_0,y_0)$ we thus should be able to write the field in the form 
$T \cong u(t) (x-x_0) + v(t) (y-y_0)$ with $u$ and $v$ complex.  Therefore to study analytically the 
dynamics of the field near the core of the vortex we make an ansatz of the type:
\be
\label{vortexAnsatz1}
T_1\left(t, x, y\right) = u\left(t\right) (x - x_0),\qquad  
T_2\left(t, x, y\right) = u\left(t\right) (y-y_0).
\ee
We have chosen $v(t) = i \, u(t)$ and $u(t)$ real since the vortex solution is expected to take the form
\be
\label{vortexAnsatz2}
T\left(t,z,\overline{z}\right) = u\left(t\right)\prod_{i}\left(z-z_i\right) 
\prod_{j}\left(\overline{z}-\overline{z}_i\right)
\ee
where we have defined complex coordinates $z = x + i y$, $\bar{z} = x - i y$.
The profile (\ref{vortexAnsatz2}), with $u(t) = \mathrm{const} \rightarrow \infty$ was used for the 
multi-vortex solutions of
\cite{Tye:ImprovedAction} where it was shown that each holomorphic zero of (\ref{vortexAnsatz2}) corresponds
to a brane and each anti-holomorphic zero of (\ref{vortexAnsatz2}) corresponds to an antibrane.
We insert now the ansatz (\ref{vortexAnsatz1}) into the equations of motion (\ref{vortexEOM}) which 
corresponds to studying the dynamics close to the core of a single vortex located at $z_0 = x_0 + i y_0$.
In this regime the equations of motion for the real and imaginary parts of the field, $T_1$ and $T_2$, 
give the same equation for the slope near $z=z_0$:
\begin{equation}
\label{vortexSlope}
  \ddot{u} = \frac{2}{b^2}\, u\, ( 1 + u^2 ) + \frac{3\, u\, \dot{u}^2}{1 + 2 u^2}
\end{equation}
where we have taken $H=0$ for simplicity.  We have verified numerically that this equation yields solutions 
which diverge in a finite time for generic initial data.  In the regime where $u(t)$ and its derivatives
are large the dominant contribution to (\ref{vortexSlope}) is
\begin{equation}
  \ddot{u} \cong \frac{2 u^3}{b^2} + \frac{3}{2}\frac{\dot{u}^2}{u}
\end{equation}
which has the solution 
\be
u(t) = \frac{b}{2(t_c-t)}.
\ee
where the critical time when the slope diverges, $t_c$, is fixed by the initial data.  This singular 
behavior corresponds to the finite-time formation of a codimension-two brane in the annihilation of a 
brane-antibrane pair.

\subsection{Alternative Complex Tachyon Action}

The action (\ref{vortexaction}) used above has the advantage of making the analysis tractable,
and the resulting dynamics is analogous to kink formation.  However, as discussed in subsection 
\ref{vortexActionSection} this action has theoretical drawbacks.  Here we consider an alternative description
of the complex tachyon dynamics.

The tachyon action has been calculation in boundary string field theory
(BSFT) in \cite{BSFT} by assuming a linear tachyon profile.  For a linear profile, gauge and space-time 
transformations allow one to write $T = u_1 x + i \, u_2 y$, and the action one derives is
\begin{equation}
\label{BSFTaction}
  S = -2 \tau_p \int d^{\,p+1} x \, \exp \left[ - 2 \pi \alpha' \left( (u_1 x)^2 + (u_2 y)^2 \right)\right] 
                F(4 \pi \alpha'^2 u_1^2)\, F(4 \pi \alpha'^2 u_2^2)
\end{equation}
where the function $F(z)$ is given by
\[
  F(z) = \frac{4^z z \Gamma(z)^2}{2 \Gamma(2 z)}.
\]

To make our analysis tractable we generalize the action (\ref{BSFTaction}) to nonlinear profiles using two 
simplifications.  The first is to consider a generalization which reduces to 
(\ref{BSFTaction}) for linear tachyon profiles only when $u_1 = u_2 = u$.  We feel this is a reasonable 
simplification since for the profile $T = u_1 x + i \, u_2 y$ to minimize the action one requires both $u_1$
and $u_2$ to be infinite.

Our next simplification is to replace the complicated function $F(z)$ by $\sqrt{1 + \pi z}$.  As 
justification we note that these two functions have identical behavior at large $z$ since 
$F(z) \rightarrow \sqrt{\pi z}$ for $z \rightarrow \infty$.

We consider, then, the action
\begin{equation}
\label{improvedaction}
  S = - 2 \tau_p \int \sqrt{-g} \, d^{\,p+1} x \exp \left(-|T|^2 / \,b^2 \right) 
  \left( 1 + \partial_{\mu} T \partial^{\mu}  \bar{T}  \right)
\end{equation}
where we have performed a rescaling of $T$ and, for consistency with (\ref{BSFTaction}), 
$b=\sqrt{\pi \alpha'}$.  This
action has also been studied in connection with tachyon condensation in \cite{DescentRelations}.  Writing
the tachyon field in real and imaginary components as $T(t,x,y) = 
T_1(t,x,y) + i \, T_2(t,x,y)$ the equation
of motion one derives from (\ref{improvedaction}) is
\begin{equation}
\label{improvedEOM}
  g^{\mu\nu} \nabla_{\mu} \nabla_{\nu} T_I + \frac{1}{b^2} 
  \left[ T_I \left( 1 + g^{\mu\nu} \nabla_{\mu} T_K \nabla_{\nu} T_K \right) 
  -  2 T_K g^{\mu\nu} \nabla_{\mu} T_K \nabla_{\nu} T_I \right] = 0.
\end{equation}

We have solved the system of two coupled partial differential equations (\ref{improvedEOM}) on a lattice, as
in subsection \ref{vortexNumericalSection}.  The results are similar to those following from the action 
(\ref{vortexaction}), though in this case the nonlinear effects are less dramatic.  One still has defect
formation, though on a longer time scale and defects begin to form for somewhat larger radius of 
compactification ($R \sim 30 M_T^{-1}$ is sufficient).  The qualitative result that cosmic strings can form
even for compactification scales well below the Hubble scale is unchanged.
We consider now the dynamics near the core of the defect by plugging the ansatz
\[
  T_1(t,x,y) = u(t) (x-x_0), \hspace{5mm} T_2(t,x,y) = u(t) (y-y_0)
\]
into (\ref{improvedEOM}) and working only to leading order in $(x-x_0)$, $(y-y_0)$.  For $H=0$ the 
equation for the slope of the field at the core of the kink is
\be
\ddot{u}-\frac{1}{b^2}u = 0
\ee
leading to an exponentially increasing slope. We see that in this case the slope does not become
singular in finite time, which can result in a different final density of defects.

\section{Comparison with Ordinary Cosmic String Formation}
\label{sect5}

If the potential for the tachyon has the usual runaway form, $\exp\left(-T^2/b^2\right)$, then once the field
starts rolling, it continues rolling towards $T = \pm \infty$. The gradient force is
insufficient to stop or reverse the rolling. 
In this section we remind the reader how this differs from the mechanism for formation
of defects in conventional field theories,
where their production is much more strongly suppressed.  The conventional case corresponds to
a theory with a global 
$U\left(1\right)$ symmetry, the potential $\frac{\lambda}{4}\left(\left|\phi\right|^2 - \sigma^2\right)^2$
and a standard kinetic term.  

In the case of  $\phi^4$ theory, the evolution is well-behaved and one can 
follow the formation and the annihilation of kinks and anti-kinks indefinitely into the future. Comparing 
the two potentials
$\exp\left(-T^2/b^2\right)$ and $\frac{\lambda}{4}\left(\left|\phi\right|^2 - \sigma^2\right)^2$
(see figure \ref{phi^4}),
we see that in the $\phi^4$ theory one expects oscillations of the field which can restore symmetry and wipe
out the defects.
\EPSFIGURE[ht]{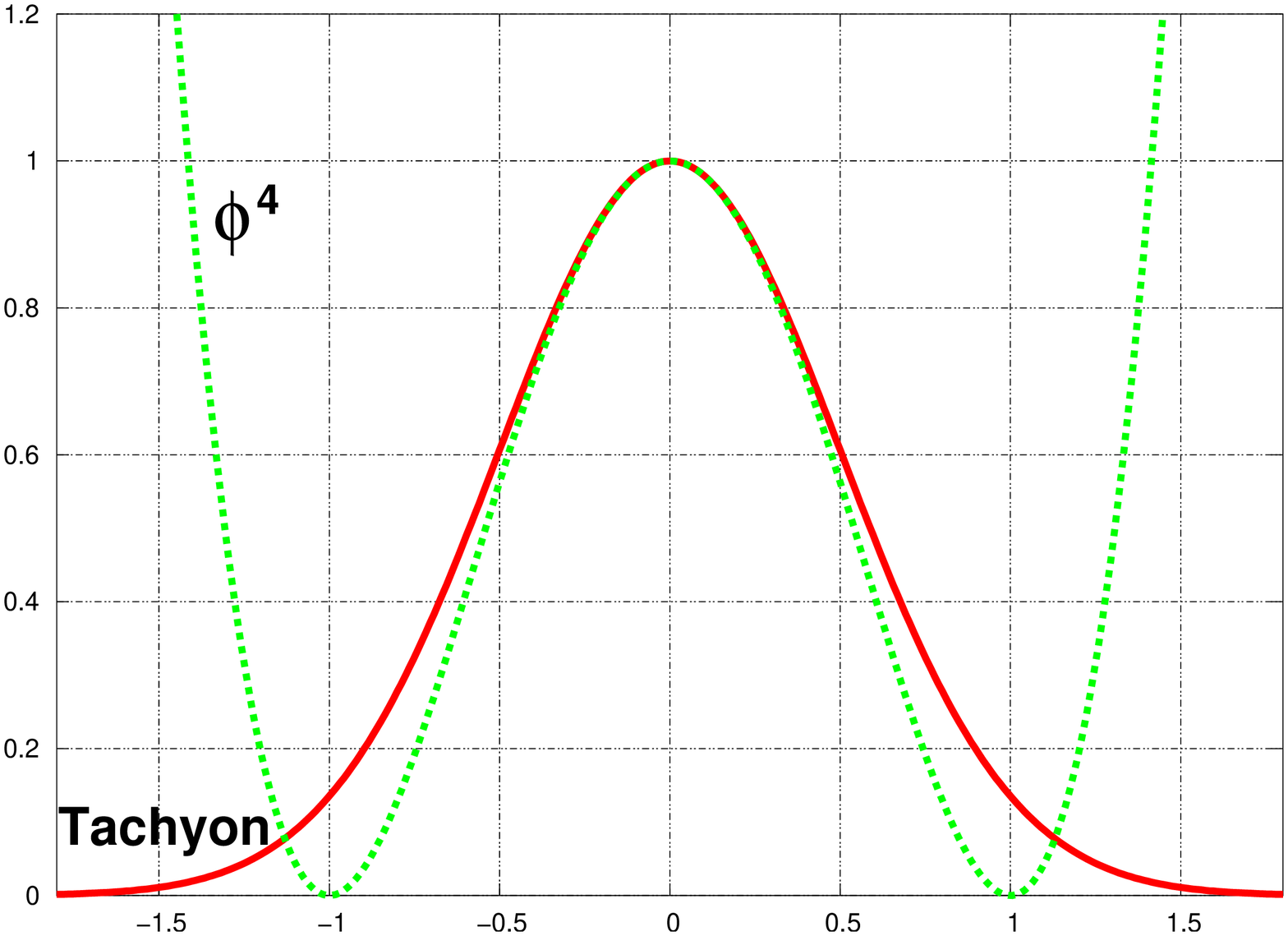,width=3.75in,height=1.95in}{Comparison of the tachyon and $\phi^4$ 
potentials. In the case of the $\phi^4$ theory, the finiteness of the slope of the kink, as well as large 
oscillations of the field, strongly suppress the formation of defects. \label{phi^4}}
\DOUBLEFIGURE[b]{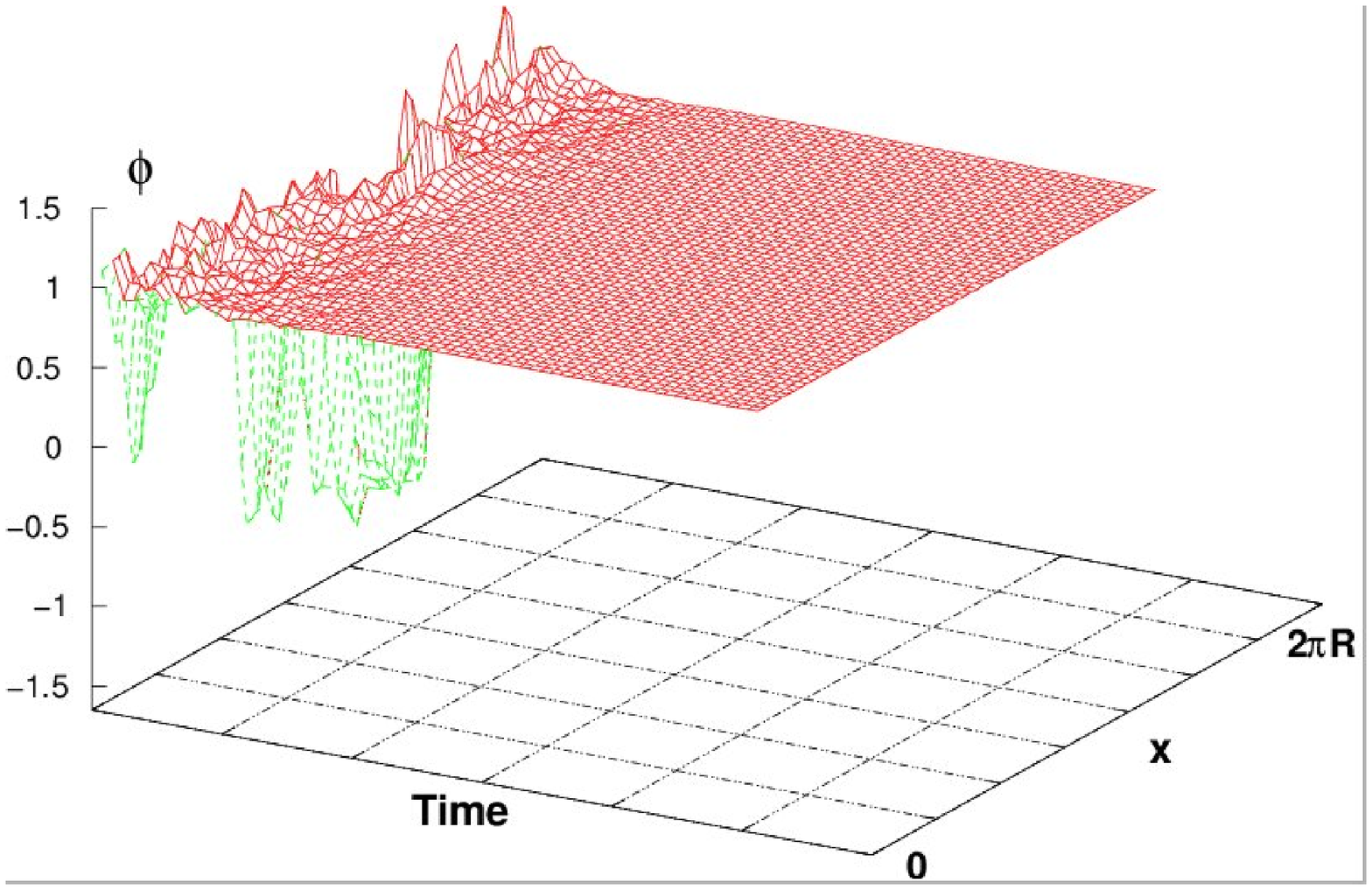,width=2.75in,height=2in}{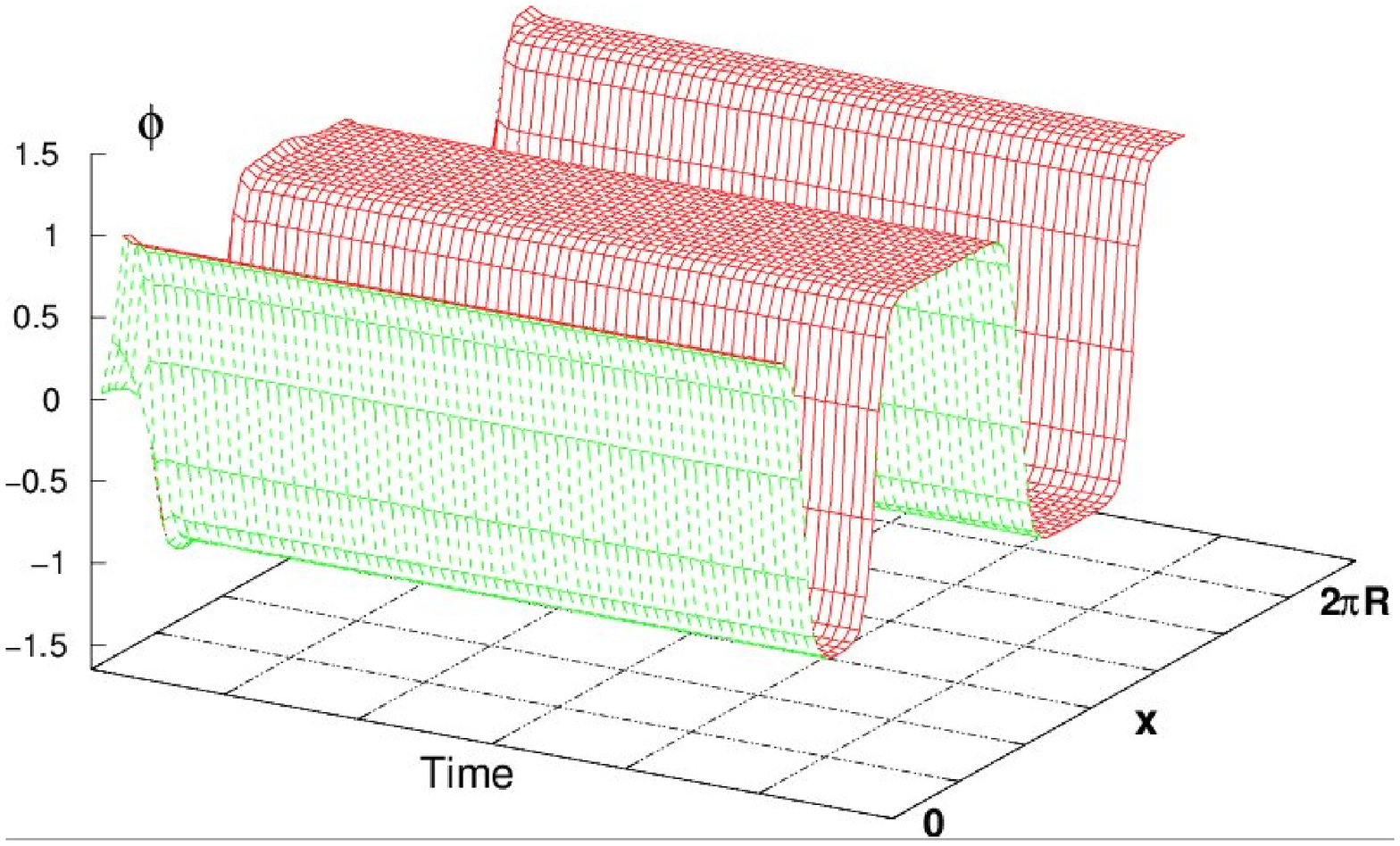,width=2.75in,height=2in}{The effect of damping on the formation of vortices in the $\phi^4$ theory. 
Small damping results in a large number of oscillations of the field, and effective
homogenization.\label{dampingFig1}}{The effect of damping on the formation of vortices in the $\phi^4$ 
theory. An unphysically large damping is used in order to show that the density of defects 
which survive is larger.
\label{dampingFig2}}

In that case the final density of defects formed  depends strongly on how fast the 
oscillations are damped, either through Hubble expansion, or through the
coupling of the field $\phi$ to other fields. In figures \ref{dampingFig1} and \ref{dampingFig2} we show the 
effect of the damping in the $\phi^4$ theory for radius of compactification $R=60 M_s^{-1}$.
A large value of $H$ results not only in a higher density of defects, but also 
slows down the motion of the defects that have formed, effectively reducing the rate at which 
they annihilate.  This is in contrast to the tachyon field theory, in which case the Hubble damping
plays essentially no role in determining the final density of defects.


\section{Cosmological Consequences}
\label{Consequences}

\subsection{String defects}
Our studies of defect formation in compact space can easily be extrapolated to imply a density of roughly
one defect per string volume in the non-compact directions aswell.
We now turn to the cosmological implications of having a very large initial 
density of strings.  In normal cosmic string networks, the details of the
initial conditions are not important because the network quickly reaches a scaling
solution.  This can be demonstrated quite simply, using the ``one-scale'' model for
the energy density $\rho$ in long strings whose characteristic length is $L$ \cite{Shellard}:
\beq
\label{osmodel}
	\dot\rho \cong -2 H \rho - f(P){\rho\over L}.
\eeq
The terms on the r.h.s.\ represent respectively the effects of dilution through
expansion and the loss due to breaking off small loops, which eventually disappear
by shrinking and emitting gravity waves.  $f(P)$ is a function of the intercommutation
probability, which is believed to go like $f\sim\sqrt{P}$ \cite{Sakel,DamourVilenkin1}.  
Taking $\rho = \mu/L^2$ (where $\mu$ is the string tension)
and $H=\beta/t$, one can verify that this has a stable attractor solution
\beq
	L = \gamma(t)\, t \equiv {f(P)\, t\over 2(1-\beta)}
\eeq
known as the scaling solution, since the string length becomes a constant fraction
of the horizon size $H^{-1}$, and the energy density in long strings also tracks
that of the dominant component,
\beq
	\rho = {\mu\over \gamma^2 t^2}.
\eeq
If the initial energy density was much greater than the scaling value, we can
find the time scale for reaching scaling by solving (\ref{osmodel}) in the 
approximation that the Hubble expansion term is negligible compared to the 
loop-emission term, giving the solution $\gamma(t) = \gamma_0 + \frac12\,f(P)\,\ln(t/t_0)$.
Inverting, we find that the time required to reach a value $\gamma= f/(2-2\beta)$,
starting from high densities where $\gamma_0\ll 1$, is 
\beq
\label{timescale}
	t \cong t_0\, e^{1/(1-\beta)}
\eeq
which is not much greater than $t_0$.  For a radiation dominated universe, with $\beta=1/2$,
the scaling solution is reached in $e^2 / 2$ Hubble times  in the usual case, and subsequent
evolution is quite insensitive to the initial conditions.  This conclusion is unchanged even
for very small intercommutation probabilities.

We have investigated the approach to the scaling solution  using a
more detailed model of network evolution, which takes into account the possibility that
loops may reconnect to long strings when the initial density is very high, and thereby
possibly
delay the onset of scaling.  Suppose that the density of small loops with characteristic
size $l$ is $1/x^3$, defining the average distance between loops at a given time.
 One can estimate that the rate per unit volume for loops to recombine with long strings is
\beq
	{dn_s\over dt} = \tilde f \, x^{-3}\, L^{-3}\, l\, L^2 \, 
	{v_{\rm rel}\over \min(x,L)}.
\eeq
Here $\tilde f$ is the probability of a reconnection, $x^{-3}$ and $L^{-3}$ are the number
densities of loops and long strings, respectively, $v_{\rm rel}$ is the relative velocity
between loops and strings, which we take to be $O(1)$, and $\min(x,L)$ is the distance a loop
typically travels before reaching a string.  The probability of a collision must be
proportional to $l$, not $l^2$, since the size of the loop in the direction parallel to the
long string does not affect the cross section. 

In this model, long strings and small loops are treated as
two separate components, $\rho_s = \mu/L^2$ and $\rho_l = \mu l/x^3$, whose energy densities are governed by
\beqa
	\dot\rho_s  &=& -2\,H \rho_s - f{\rho\over L} + \mu\, l\,{dn_r\over dt}, \nonumber\\
	\dot\rho_l &=& -3\,H \rho_l  +f{\rho\over L}  - \mu\, l\,{dn_r\over dt} -
	\Gamma G \mu^2 {1\over x^3}.
\eeqa
The last term represents the power emitted by loops due to gravitational radiation, 
$\Gamma G \mu^2$ \cite{DamourVilenkin2}, where $\Gamma \cong 50$ and $10^{-11} \lsim G \mu \lsim 10^{-6}$
\cite{Tye:StringProduction}-\cite{CosmicStrings}.  The loop size is taken to always be a fixed fraction of 
the long string correlation length: $l/L = \Gamma G \mu$, so long as this is not
smaller than the fundamental string length scale $l_s$.  We have integrated 
these equations numerically, together with the Friedmann equation and the evolution
equations for energy density in visible and gravitational radiation, keeping the short
distance cutoff $l_s$ on the size of the loops (in fact the results do not change noticeably
if we assume the loops remain as small as this cutoff).  This
more detailed study confirms that the scaling solution is attained in only a few Hubble times,
as can be seen from the time evolution of the fractions of the critical density for each component, shown in
figure \ref{omega}.  We note, however, that in the above analysis we assumed that string
velocities remain of order unity; if there is significant freezing out of the relative string
motions in the large  in the large dimensions this could have significant impact on the
approach to scaling \cite{Shellard}.  We note also that the effect of friction due to particle scattering
may also significantly alter this picture \cite{friction}.\footnote{We thank J.J. Blanco-Pillado and Carlos
Martins for bringing this to our attention.}

\EPSFIGURE[ht]{omega4.eps,width=3.75in,height=2.5in}{Fraction of critical density versus time for long 
strings, loops, and visible radiation starting from initial values $\Omega_s=0.4$, $\Omega_l = 0.1$ and 
$\Omega_{rad} = 0.5$.  Not shown is the fraction of energy density in gravity waves produced by decay
of the loops. \label{omega}}

The initial density of strings is many orders of
magnitude greater than the Kibble estimate, 
which gave the initial correlation length $L$ as $\sim
H^{-1}$ at the moment of formation of the network;  instead, the initial energy density of 
the network is comparable to the total energy density available, $\sim\mu^2$, so that 
$L\sim \mu^{-1/2}$, smaller by a factor of $\mu^{-1/2}/M_p$ than the Kibble value.  
Since $\rho = \mu/L^2$, the initial density is greater by a factor of $M_p^2/\mu$ than
the Kibble value. However,
this huge excess is so shortlived that
it has no observable effects.  For instance, the contribution from the early nonscaling 
regime to the gravity wave background is negligible due to the small size of the loops which
are formed and subsequently radiate during this era.  Following ref.\ \cite{DamourVilenkin2},
one can estimate the amplitude of these gravity waves at frequency $\omega$ as being of order
\beq
\label{strain}
	h \sim {(G\mu)^{7/6} \rho_0^{7/12}\over M_p^2 \omega^{1/3}} \sim 10^{-66}
\eeq
where $\rho_0$ is the present energy density of the universe.  The estimate (\ref{strain}) is
some 40 orders of magnitude below the sensitivity of LIGO at the frequencies of interest,
$\omega\sim 100$ Hz.  As for the cosmic microwave background, the wavelength of density
perturbations created during the nonscaling regime of the network is too short to be
relevant: initially $\lambda\sim\mu^{-1/2}$, which gets stretched to the scale of the present
energy density $\lambda\sim \rho_0^{-1/4} \sim 0.1$ mm.  This length scale is also too small
to be relevant for the formation of primordial black holes (PBH's) since the mass contained
in volume $\lambda^3$ is far below that needed for cosmologically long-lived PBH's.

\subsection{Domain Walls}
\label{DWsubsection}

Thus the fast approach to the scaling solution for three-dimensional string networks erases
all sensitivity to the initial conditions, even though the initial density is orders of
magnitude greater than for conventional cosmic strings, and this is true regardless of the reduced
intercommutation probability.  However, there are situations where our modified understanding
of the network's initial conditions may make a dramatic difference: namely, in the case of
higher dimensional defects.  Let us illustrate with the example of D5-$\overline{{\rm D}5}$
annihilation, where two of the dimensions are wrapped on an internal manifold with
coordinates $(y_1,y_2)$ and the remaining three dimensions span the euclidean space $(x^1,x^2,x^3)$.  
The codimension-two defects which form from the annihilation are D3
branes, and these can have various orientations with respect to the world-volume of the
parent branes.  The choices are exemplified by the three situations:
\begin{enumerate}
\item extended in $x_3,y_1,y_2$ directions,  localized in $x_1,x_2$: looks like D1 in 3D. 
\item extended in $x_2,x_3,y_2$ directions,  localized in $x_1,y_1$: looks like D2 in 3D.
\item extended in $x_1,x_2,x_3$ directions,  localized in $y_1,y_2$: looks like D3 in 3D.
\end{enumerate}

The first case looks like ordinary cosmic strings  to the 3D observer since the  defects have
only one long direction among the three large ones.  Their effects have already been
discussed.  Case 3 is a network of 3-branes, all of whose dimensions are large.  Their 
tension will contribute to the effective 3D cosmological constant.\footnote{These can be safely assumed to annihilate quickly since they are not separated in the expanding
space $(x^1,x^2,x^3)$.}\ \  Case 2, illustrated in
fig.\ \ref{dwall}, is the interesting one because these appear as domain walls to the
3D observer, and their energy density redshifts too slowly: $\rho\sim 1/a^2$ in terms of the
scale factor of the large dimensions. A single domain wall of tension $\tau_2 = \eta^3$
within our horizon would dominate the present energy density unless $\eta \lsim 1$ MeV. 
The effective tension of a 3-brane wrapping one compact dimension of size $R$ is $\tau_2 =
R\tau_3\sim \mu^{3/2}$; hence the string scale would have to be $\lsim 1$ MeV, absurdly
small.

\EPSFIGURE[ht]{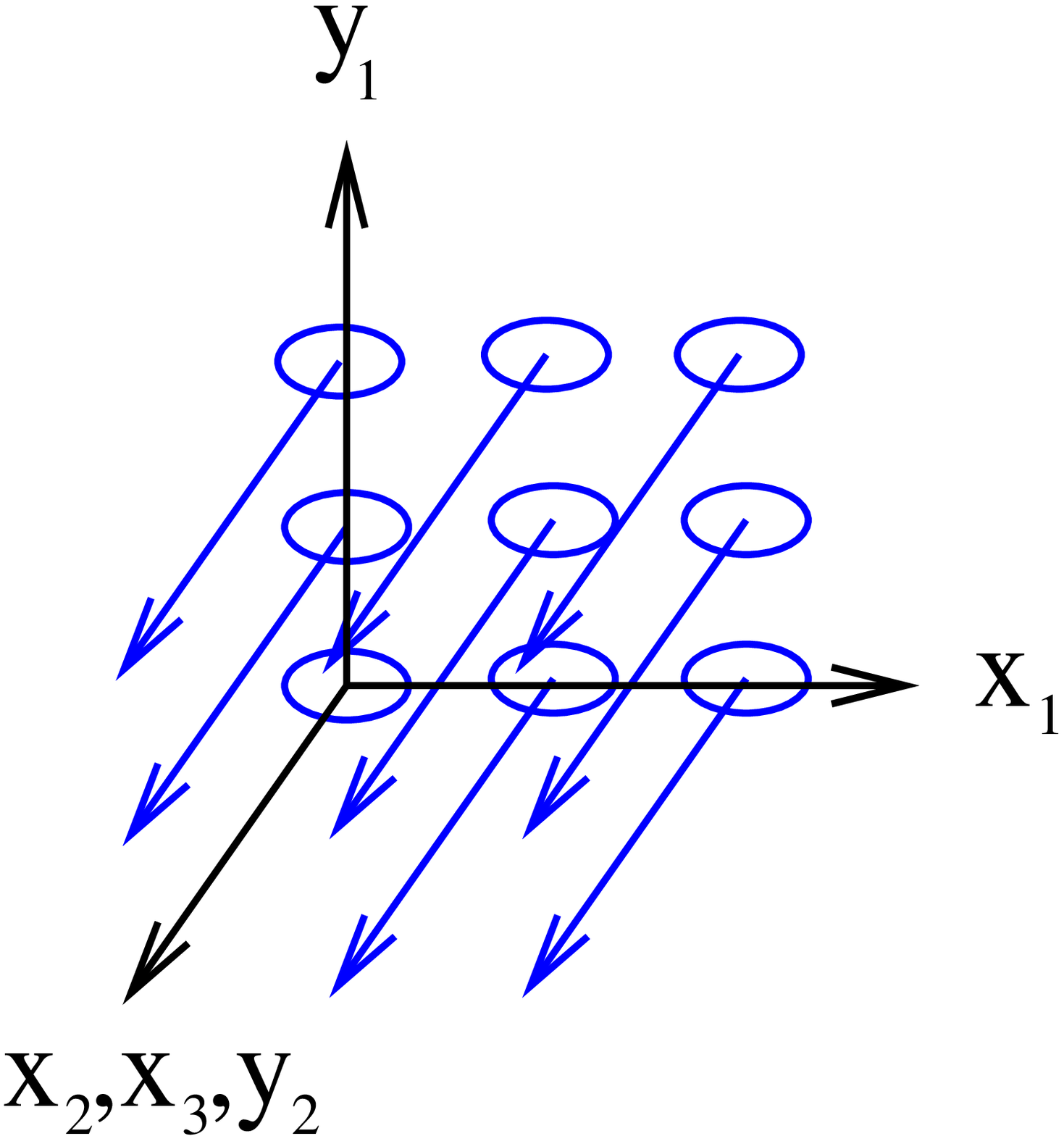,width=2.5in}{An array of codimension-two D3-branes from D5-anti-D5 
annihilation, partially localized in the compact dimensions, which look like strings in the $x_1$-$y_1$ 
plane, and domain walls in the large $x_i$ dimensions.\label{dwall}}

Our conclusion differs from that of ref.\ \cite{JST}, which speculated that no defects
partially localized in compact directions can form.  The argument was based on the Kibble
mechanism, and thus assumed the correlation length could not be smaller than $H^{-1}\sim
M_p/\mu$,  whereas the size of the compact dimensions must be much smaller, of order
$\mu^{-1/2}$.  We have seen that in fact the initial correlation length is of the same order 
as the string scale, so that this argument is invalidated.

Still, one might be skeptical as to whether defects partially localized in the compact
directions can survive until today, since intuitively they might be able to find each other
and annihilate very quickly in the compact space.  Ref.\ \cite{MajumDavis} attempts to address
this question, and concludes that domain wall defects like those in case 2 {\it cannot}
efficiently annihilate, since they are not completely localized in the compact (nonexpanding) space.
The analysis of \cite{MajumDavis} assumes that the
number density of D$p$-branes satisfies a rate equation which is nearly the same as 
that governing monopoles:
\beq
\label{wrongeq}
	\dot n = -(3-p_{||}) H n - {\Gamma\over T^{D-2}} n^2
\eeq
where $p_{||}$ is the number of dimensions of the brane spanning the large dimensions, and
$D$ is the total number of spacetime dimensions.   This ignores the effect of
self-intersections for reducing the density of long defects, which is known to be the
dominant means for string networks to reach scaling (cf.\ eq.\ (\ref{osmodel})).  Further,
it unrealistically assumes that the defects are parallel, so that they will annihilate rather
than intercommute when they meet.  It is therefore not immediately clear how far we can trust their
conclusions \cite{MajumDavis}.

On the other hand, numerical evolution of domain wall networks shows that self-intersections
are not generic, and it is suggested that the dominant energy-loss mechanism is direct
gravitational radiation rather than through collisions  \cite{GarHind}.  Furthermore it is observed
that the approach to the scaling solution is slower for domain walls than for cosmic strings
\cite{OMA}.  

\subsection{Full String Network Simulations}

To attempt to address the issue of whether domain walls disappear or not, we have considered
 the dynamics of 
D$3$-branes in $(5+1)$ dimensions in the approximation of projecting out the dynamics in one compact 
direction, $y^2$, and one noncompact direction, $x^3$, to give an effective $3+1$-dimensional system.  In 
this case the D$3$-branes appear as one-dimensional objects (see figure \ref{dwall}) and the dynamics can be
modeled by considering string evolution in an anisotropic space with two large and one small dimension.  In 
this setup those ``strings''---string-like from the $(x^1,x^2,y^1)$ point of view---which span the 
large dimensions $x^1$ and $x^2$ appear as domain walls to the $3$D observer while the
 ``strings'' which 
span the compact dimension $y^1$ appear as cosmic strings in $3$D.

Following the setup of Smith and Vilenkin \cite{vilsmith} we performed numerical simulations of the
defect evolution in this approximation, keeping track of the extent to which defects preferentially
spanned the compact direction $y^1$---thus appearing as strings in the $x^1, x^2, x^3$
subspace, relative to spanning the
large directions, which appear as domain walls.  Figure \ref{fig:simulate} shows the fractional energy 
density in wound strings  wrapped about the three directions $x^1, x^2, y^1$ as a function of time. This
particular run started with equal energy in each direction so that the correlation lengths, and hence the 
interaction rates, were roughly the same. We observe that the branes wrapping the large directions
lose their winding energy more quickly than those wrapping the compact direction:  this can be attributed to 
several factors, including the smaller cross-sectional area of the long strings, and the greater energy
radiated away when two long strings annihilate.  The implication is that defects which look
like strings to the 3D observer tend to survive preferentially over those that appear as
domain walls.  However since this is a toy model for the actual higher-dimensional defects,
such a conclusion awaits validation from actual domain wall simulations.

\EPSFIGURE[ht]{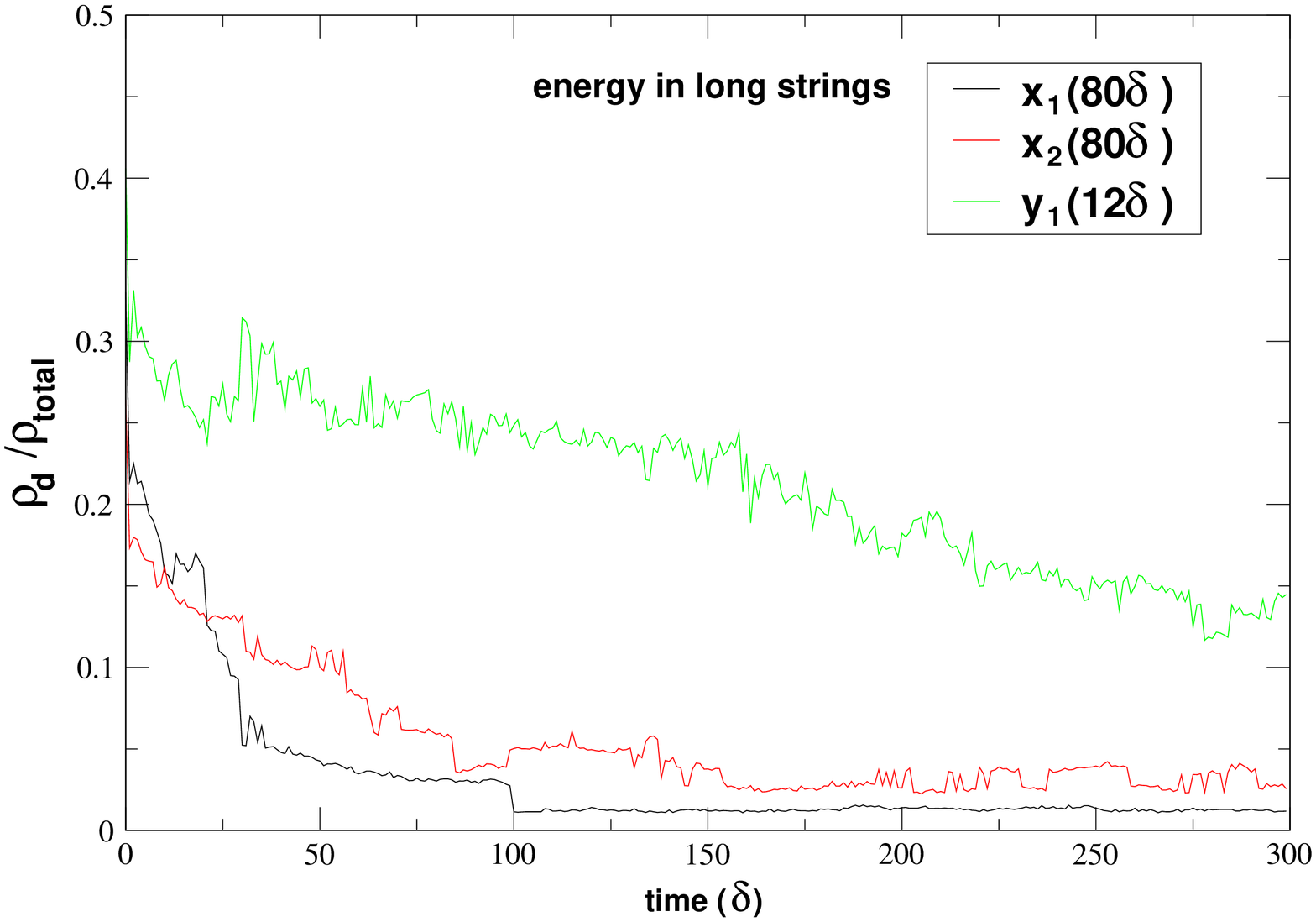,width=5in,angle=0}{Fraction of wound string density about the three 
anisotropic directions (two large and one small) starting from equal energy density in strings wrapping each
direction.  Top curve is for strings winding in the small direction.\label{fig:simulate}}

As further evidence that defect evolution after formation tends to favor the defects which are localized
in the large dimensions (and hence tends to favor the survival of cosmic string-like defects) we consider
a field theory simulation of domain wall evolution for a real scalar field with potential 
$\frac{\lambda}{4}\left(\phi^2 - \sigma^2\right)^2$ and standard kinetic term 
\footnote{Though it is of more direct interest to consider the dynamics of the tachyon field theory with 
action (\ref{kinkaction}) the finite time slope divergence prevents us from following the dynamics after the
defects have formed.}\ \   We studied the formation and evolution of domain walls in this theory in an 
anisotropic $(3+1)$-dimensional space with two large and one compact dimension.  Starting from a
random initial profile close to the false vacuum (as in subsection \ref{kinkNumerical}) we find that the
evolution of the defects after formation tends to annihilate domain walls which are localized in the compact
direction and to favor the survival of domain walls which are localized in the large directions.  Figure
\ref{DWfieldsim} shows a plot of the domain wall network late in the evolution after the domain walls 
localized in the compact direction have disappeared.  

We also explored the effect of increasing the
anisotropy, varying the initial distribution of winding modes, and varying
the intercommutation probability; however, the system consistently evolved to
favor winding about the compact direction. These results seem to 
corroborate the claim that domain wall-like defects are suppressed.

\EPSFIGURE[ht]{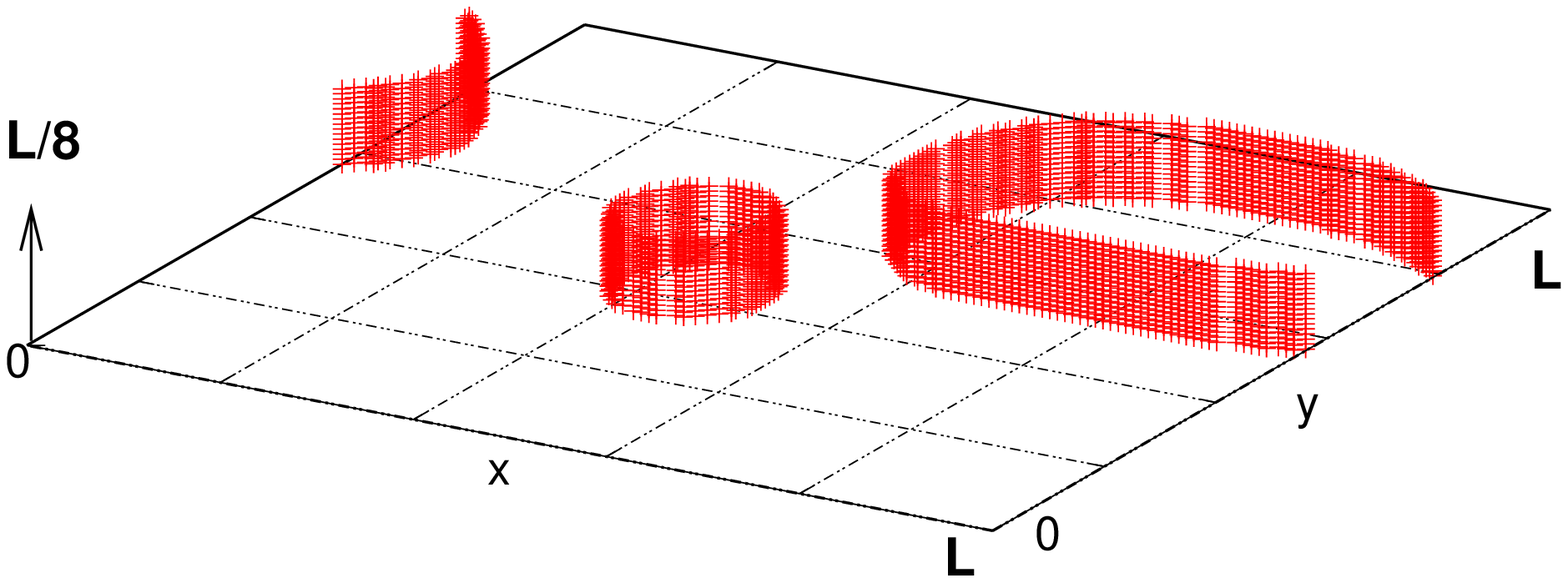,width=5in,height=3in,angle=360}{Plot of domain wall network in an 
anisotropic
space with two large and one small dimensions with $L = 80 M_s^{-1}$.  This snapshot is taken late in the 
evolution and shows that domain walls which are localized in the large dimensions are preferred by the 
defect evolution.\label{DWfieldsim}}

We note that there are several factors which we have not taken into account which may radically
alter the picture.  For example, in figure \ref{fig:simulate} we have unrealistically neglected the dynamics
in two dimensions to facilitate numerical studies.  Furthermore, in both of the examples above we have 
neglected the expansion of the large dimensions---an effect which could play an important role in 
the dynamics.  It is therefore unclear whether domain walls will pose a cosmological problem in models of 
brane-antibrane inflation where the branes driving inflation wrap the compact space.  We feel that this is 
a problem which merits further investigation and it is likely that a complete resolution of this issue
will require full simulations of $3$-brane dynamics in an anisotropic $(5+1)$-dimensional spacetime
with $3$ expanding dimensions.  While such simulations would be of interest, they are beyond the scope of 
the present work.

\subsection{Monopoles}

Finally consider an example of how monopole-like defects may be formed through a cascade of annihilations in 
D5-$\overline{{\rm D}5}$ inflation.  The initial state D5 and $\overline{{\rm D}5}$ span the three large 
dimensions and wrap two compact dimensions.  These may produce D3 and $\overline{{\rm D}3}$ which wrap the
two compact dimensions and are extended in one large dimension; 
hence these defects appear string-like
to the $3$-dimensional observer.  A parellel D3-$\overline{{\rm D}3}$ pair may then annihilate to produce 
D1 branes and antibranes which can span the same large dimension as the parent 
D3-$\overline{{\rm D}3}$, or alternatively could
wrap the compact dimensions.  Those D1-branes which span the large dimension appear as cosmic string 
defects to the $3$-dimensional observer while those which wrap the compact dimensions will 
appear as point-like (monopole) defects.  If the compact dimensions admit nontrivial 1-cycles (like $T^2$ for example) then
these monopoles will be stable.  Our results indicate that in general both string-like and monopole-like 
defects should be produced in this cascade.  
To understand the subsequent evolution of such defects we
consider numerically the dynamics of D$1$-branes in a $3$D space with one large dimension and two small 
dimensions.  Figure \ref{fig:simulate2} shows the fractional energy 
density in wound strings  wrapped about the three directions $x^1, y^1, y^2$ as a function of time. 
Strings wrapping the large direction $x^1$ appear as genuine cosmic strings to the $3$D observer while
strings wrapping small directions $y^1,y^2$ appear as monopole-like defects in $3$D.  As in
subsection \ref{DWsubsection} we start with equal energy in each direction so that the correlation lengths 
are roughly equal.  Again we find that  strings wrapping the large direction lose their winding energy 
quicker than  those wrapping the compact directions.  Physically, this suggests that monopoles are 
preferentially produced over cosmic strings in this particular cascade of annihilations.

\EPSFIGURE[ht]{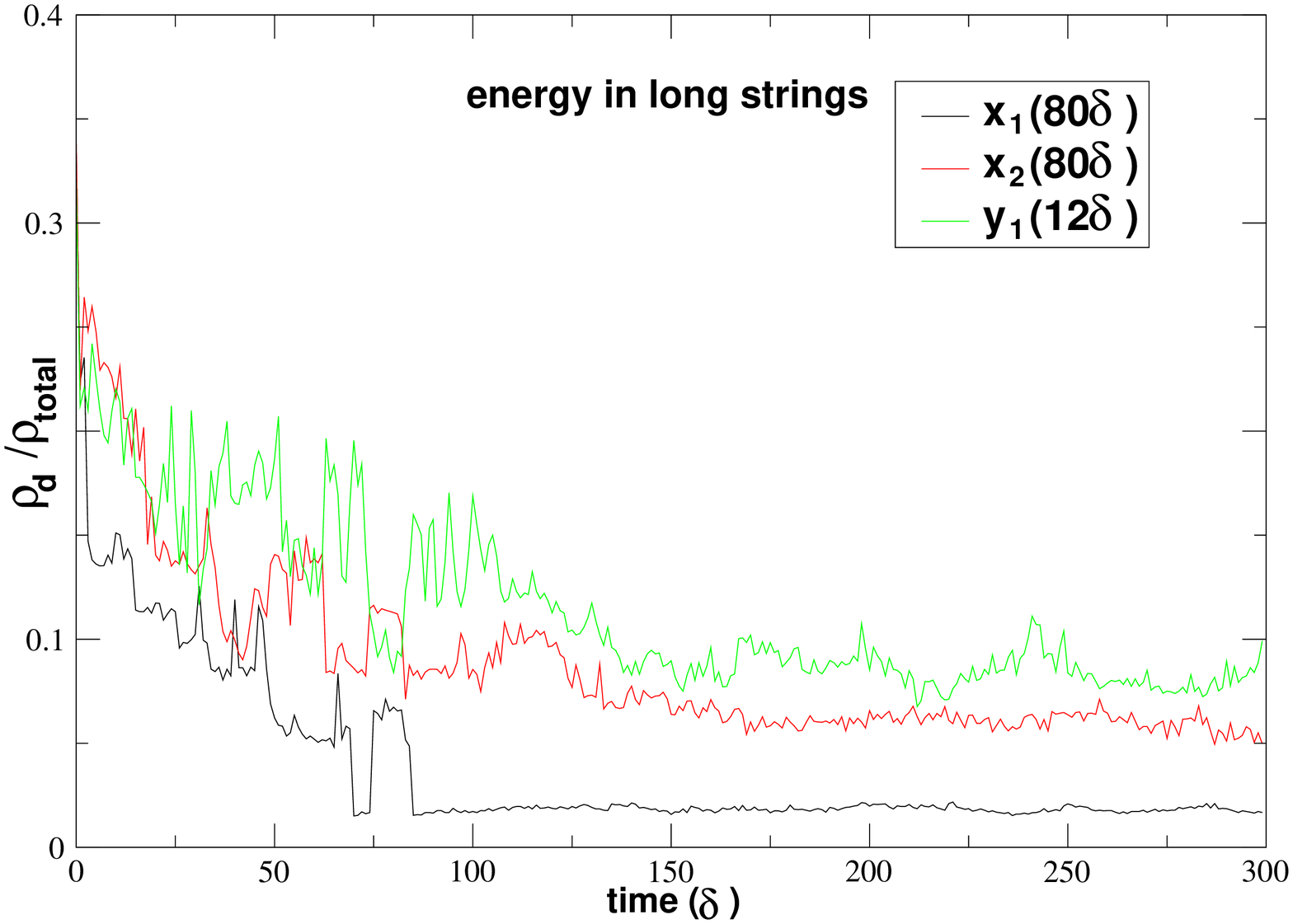,width=5in,angle=0}{Fraction of wound string density about the 
three anisotropic directions (one large and two small) starting from equal energy density in strings 
wrapping each direction.\label{fig:simulate2}}

The question of whether such monopole-like defects will pose a 
cosmological problem depends crucially on the long-range forces between these defects and is, we feel, an 
issue which merits further investigation (see \cite{SF} for a solution of the monopole problem which is 
independent of inflation).

\section{Summary and Conclusions}
\label{Conclusions}

We have studied the formation of topological defects during the decay of a nonBPS brane or a
coincident brane-antibrane pair. The problem was treated both analytically, by solving the 
equation of motion for the tachyon field at the core of the defect, as well as numerically,
by evolving
the tachyon field on a lattice. We showed that defects form with a correlation length proportional to the 
string length rather than the horizon. Defects localized within compact dimensions can form even if the
compactification radius is as small as $7M_S^{-1}$ and the tachyon dynamics is insentitive the the Hubble
damping. Depending on the exact form of the action ({\it e.g.}, Sen's version, or the boundary string field
theory version) the slope of the field at the defect could either increase exponentially  in time, or else 
diverge within a finite time, potentially changing the initial density of defects. We compared the evolution
of the tachyon field to that of a scalar field in the $\phi^4$ theory and noted that the most
efficient way to suppress the formation of defects is through symmetry restoration, caused by
large oscillations of the scalar field. This is not possible if the potential has a runaway
form, as for the string tachyon, which inevitably leads to the formation of a higher density
of defects in the string case. Once the defects form the field theory description is no
longer adequate, so in order to analyze the annihilation of the defects formed one has to 
use a description in terms of branes and antibranes interacting in a compact space.

As a result, the initial density of string defects is much greater than previous estimates.
For strings which are genuine 1D objects, we showed that the string network nevertheless attains
scaling behavior within just a few Hubble times, so that there are no observable 
consequences of the initial high string density.  Of course this assumes that the network is
not frustrated \cite{frustrated}, so that scaling can indeed be achieved.  Whether this is
the case for cosmic superstrings which are bound states of fundamental and D-strings is
an interesting open question \cite{polchinski-review}.

On the other hand, we argue that in models where inflation is driven by branes which wrap the compact
manifold (for example D5-$\overline{{\rm D}5}$), domain wall-like and monopole-like defects are inevitably 
produced.  The stability and subsequent evolution of such defects is complicated and may depend crucially
on the details of the compactification, for example, on whether or not the compact manifold which the parent 
branes wrap admits nontrivial 1-cycles.  We leave detailed studies of whether such models are 
phenomenologically viable to future investigations; however, we have shown that the formation of defects at 
the end of brane-antibrane inflation is much more complicated and model-dependent than one 
might have expected. 

\section{Acknowledgments}

We would like to thank Robert Brandenberger, J.J. Blanco-Pillado, Guy Moore and Henry Tye for discussions. 
This work was supported in part by NSERC and FQRNT. 

\renewcommand{\theequation}{A-\arabic{equation}}
\setcounter{equation}{0}  

\section*{APPENDIX: Initial Conditions for Defect Formation}

Here we briefly discuss the initial conditions for defect formation at the end
of brane inflation. During the slow roll inflationary phase the tachyon behaves
as an ordinary massive Klein-Gordon scalar field  (provided $T \ll l_s^{-1}$). 
We consider here for simplicity a standard field theory of a scalar $\chi$ in 
$3+1$-dimensions whose mass squared parameter abruptly becomes negative.  This
type of theory has been considered in detail in
\cite{TachyonicPreheating1,TachyonicPreheating2}.

During the de Sitter phase (before the mass parameter becomes 
tachyonic) vacuum fluctuations yield a blackbody spectrum of produced particles  $\langle 0| N_k |0\rangle 
= ( e^{\omega_k / T} - 1)^{-1}$
with temperature $T = H / (2 \pi)$ (see, for example, 
\cite{Brandenberger2} for a review). 
However, in any region of de Sitter space which is small compared
to the Hubble scale, the space is locally Minkowski, and even in the vacuum state
there are quantum fluctuations quantified by the two point functions of the fields
\begin{eqnarray*}
\langle \chi^{\star} (\vec{k}) \chi(\vec{k}')  \rangle  &=& \frac{1}{2 |\vec{k}|} (2 \pi)^3 \delta^3 (\vec{k}-\vec{k}') \\
\langle \Pi^{\star} (\vec{k}) \Pi(\vec{k}')  \rangle  &=& \frac{|\vec{k}|}{2} (2 \pi)^3 \delta^3 (\vec{k}-\vec{k}')
\end{eqnarray*}
where $\Pi = \dot{\chi}$.\footnote{Similar initial conditions are taken for defect formation at the 
end of inflation in \cite{Kawasaki}.}
The initial stages of the string tachyon condensation are identical to the 
tachyonic preheating scenario described in \cite{TachyonicPreheating1,TachyonicPreheating2}.  The tachyonic
instability amplifies exponentially those modes with $|\vec{k}| < m$ where the $\chi$ field has mass squared
parameter $-m^2$ and the variance of the fluctuations grows as \cite{TachyonicPreheating2}
\[
  \langle  \chi^2(t) \rangle    = \langle \chi^2(0)\rangle  + 
  \frac{1}{8 \pi} \int_0^{m^2} dk^2 \frac{m^2}{m^2-k^2} \sinh^2 \left( t \sqrt{m^2-k^2} \right)
\]
where $\langle\chi^2(0)\rangle$ is a divergent vacuum contribution. The above result was
derived in (3+1) dimensions, but the generalization to higher dimensions must have the form
\[
  \langle  \chi^2(t) \rangle    = \langle \chi^2(0)\rangle  + 
  c \sum_i \int_0^{m^2} dk^2 \frac{m^2}{m^2-k^2-m^2_i} \sinh^2 \left( t \sqrt{m^2-k^2-m^2_i} \right)
\]
where $m_i$ are the masses of the Kaluza-Klein excitations.  
 These fluctuations grow to be of order the classical VEV, 
$\langle\chi^2(t)\rangle - \langle\chi^2(0)\rangle \sim m^2/\lambda$ before 
the linear treatment breaks down.  Notice that this growth
occurs on a microscopic time scale.  Since in the case of the string theory tachyon
the potential has no minimum,
we can conservatively take the initial exponential growth to be over a shorter time scale,
$\langle\chi^2\rangle \sim m^2 \sim M_s^2$.   These fluctuations are
much larger than the de Sitter fluctuations $\langle\chi^2\rangle \sim H^2$ and are thus
the dominant seeds for defect 
formation.  Furthermore, these fluctuations have a minimum wavelength comparable to the string scale which
is sufficient to initiate the formation of defects localized in the compact dimensions.  
This justifies our choice of random initial conditions for the tachyon field in the
numerical studies of defect formation.


\begin{thebibliography}{99}

\bibitem{Tye:StringProduction}

S.~Sarangi and S.~H.~H.~Tye,
``Cosmic string production towards the end of brane inflation,''
Phys.\ Lett.\ B {\bf 536}, 185 (2002)
[arXiv:hep-th/0204074].

\bibitem{JST}
N.~T.~Jones, H.~Stoica and S.~H.~H.~Tye,
``The production, spectrum and evolution of cosmic strings in brane
inflation,''
Phys.\ Lett.\ B {\bf 563}, 6 (2003)
[arXiv:hep-th/0303269].

\bibitem{CosmicStrings}

L.~Pogosian, S.~H.~H.~Tye, I.~Wasserman and M.~Wyman,
``Observational constraints on cosmic string production during brane
inflation,''
Phys.\ Rev.\ D {\bf 68}, 023506 (2003)
[arXiv:hep-th/0304188].

G.~Dvali, R.~Kallosh and A.~Van Proeyen,
``D-term strings,''
JHEP {\bf 0401}, 035 (2004)
[arXiv:hep-th/0312005].

G.~Dvali and A.~Vilenkin,
``Formation and evolution of cosmic D-strings,''
JCAP {\bf 0403}, 010 (2004)
[arXiv:hep-th/0312007].

E.~J.~Copeland, R.~C.~Myers and J.~Polchinski,
``Cosmic F- and D-strings,''
JHEP {\bf 0406}, 013 (2004)
[arXiv:hep-th/0312067].

E.~Halyo,
``Cosmic D-term strings as wrapped D3 branes,''
JHEP {\bf 0403}, 047 (2004)
[arXiv:hep-th/0312268].

L.~Leblond and S.~H.~H.~Tye,
``Stability of D1-strings inside a D3-brane,''
JHEP {\bf 0403}, 055 (2004)
[arXiv:hep-th/0402072].

K.~Dasgupta, J.~P.~Hsu, R.~Kallosh, A.~Linde and M.~Zagermann,
``D3/D7 brane inflation and semilocal strings,''
JHEP {\bf 0408}, 030 (2004)
[arXiv:hep-th/0405247].

\bibitem{GW}

T.~Damour and A.~Vilenkin,
``Gravitational radiation from cosmic (super)strings: Bursts, stochastic
background, and observational windows,''
arXiv:hep-th/0410222.

\bibitem{SS}

G.~D.~Starkman and D.~B.~Stojkovic,
``How Frustrated Strings Would Pull the Black Holes from the Centers of
Galaxies,''
Phys.\ Rev.\ D {\bf 63}, 045008 (2001)
[arXiv:astro-ph/0010563].

\bibitem{JJP}
M.~G.~Jackson, N.~T.~Jones and J.~Polchinski,
``Collisions of cosmic F- and D-strings,''
arXiv:hep-th/0405229.

\bibitem{Kibble}

T.~W.~B.~Kibble,
``Topology Of Cosmic Domains And Strings,''
J.\ Phys.\ A {\bf 9}, 1387 (1976).

T.~W.~B.~Kibble,
``Some Implications Of A Cosmological Phase Transition,''
Phys.\ Rept.\  {\bf 67}, 183 (1980).

\bibitem{Brandenberger}

J.~Magueijo and R.~H.~Brandenberger,
``Cosmic defects and cosmology,''
arXiv:astro-ph/0002030.

\bibitem{Sen:TachyonMatter}

A.~Sen,
``Field theory of tachyon matter,''
Mod.\ Phys.\ Lett.\ A {\bf 17}, 1797 (2002)
[arXiv:hep-th/0204143].

\bibitem{BraneInflation}

G.~R.~Dvali and S.~H.~H.~Tye,
``Brane inflation,''
Phys.\ Lett.\ B {\bf 450}, 72 (1999)
[arXiv:hep-ph/9812483].

S.~H.~S.~Alexander,
``Inflation from D- anti-D brane annihilation,''
Phys.\ Rev.\ D {\bf 65}, 023507 (2002)
[arXiv:hep-th/0105032].

A.~Mazumdar, S.~Panda and A.~Perez-Lorenzana,
``Assisted inflation via tachyon condensation,''
Nucl.\ Phys.\ B {\bf 614}, 101 (2001)
[arXiv:hep-ph/0107058].

C.~P.~Burgess, M.~Majumdar, D.~Nolte, F.~Quevedo, G.~Rajesh and R.~J.~Zhang,
``The inflationary brane-antibrane universe,''
JHEP {\bf 0107}, 047 (2001)
[arXiv:hep-th/0105204].

G.~R.~Dvali, Q.~Shafi and S.~Solganik,
``D-brane inflation,''
arXiv:hep-th/0105203.

C.~Herdeiro, S.~Hirano and R.~Kallosh,
late``String theory and hybrid inflation / acceleration,''
JHEP {\bf 0112}, 027 (2001)
[arXiv:hep-th/0110271].

B.~s.~Kyae and Q.~Shafi,
``Branes and inflationary cosmology,''
Phys.\ Lett.\ B {\bf 526}, 379 (2002)
[arXiv:hep-ph/0111101].

C.~P.~Burgess, P.~Martineau, F.~Quevedo, G.~Rajesh and R.~J.~Zhang,
``Brane antibrane inflation in orbifold and orientifold models,''
JHEP {\bf 0203}, 052 (2002)
[arXiv:hep-th/0111025].

J.~Garcia-Bellido, R.~Rabadan and F.~Zamora,
``Inflationary scenarios from branes at angles,''
JHEP {\bf 0201}, 036 (2002)
[arXiv:hep-th/0112147].

R.~Blumenhagen, B.~Kors, D.~Lust and T.~Ott,
``Hybrid inflation in intersecting brane worlds,''
Nucl.\ Phys.\ B {\bf 641}, 235 (2002)
[arXiv:hep-th/0202124].

K.~Dasgupta, C.~Herdeiro, S.~Hirano and R.~Kallosh,
``D3/D7 inflationary model and M-theory,''
Phys.\ Rev.\ D {\bf 65}, 126002 (2002)
[arXiv:hep-th/0203019].

N.~Jones, H.~Stoica and S.~H.~H.~Tye,
``Brane interaction as the origin of inflation,''
JHEP {\bf 0207}, 051 (2002)
[arXiv:hep-th/0203163].

M.~Gomez-Reino and I.~Zavala,
``Recombination of intersecting D-branes and cosmological inflation,''
JHEP {\bf 0209}, 020 (2002)
[arXiv:hep-th/0207278].

S.~Kachru, R.~Kallosh, A.~Linde and S.~P.~Trivedi,
``De Sitter vacua in string theory,''
Phys.\ Rev.\ D {\bf 68}, 046005 (2003)
[arXiv:hep-th/0301240];

S.~Kachru, R.~Kallosh, A.~Linde, J.~Maldacena, L.~McAllister and S.~P.~Trivedi,
``Towards inflation in string theory,''
JCAP {\bf 0310}, 013 (2003)
[arXiv:hep-th/0308055].

J.~P.~Hsu, R.~Kallosh and S.~Prokushkin,
``On brane inflation with volume stabilization,''
JCAP {\bf 0312}, 009 (2003)
[arXiv:hep-th/0311077].

H.~Firouzjahi and S.~H.~H.~Tye,
``Closer towards inflation in string theory,''
arXiv:hep-th/0312020.

E.~Halyo,
``Inflation on fractional branes: D-brane inflation as D-term inflation,''
JHEP {\bf 0407}, 080 (2004)
[arXiv:hep-th/0312042].

E.~Halyo,
``D-brane inflation on conifolds,''
arXiv:hep-th/0402155.

C.~P.~Burgess, J.~M.~Cline, H.~Stoica and F.~Quevedo,
``Inflation in realistic D-brane models,''
arXiv:hep-th/0403119.

A.~Buchel and A.~Ghodsi,
``Braneworld inflation,''
arXiv:hep-th/0404151.

\bibitem{CarollianContraction}

G.~Gibbons, K.~Hashimoto and P.~Yi,
``Tachyon condensates, Carrollian contraction of Lorentz group, and
fundamental strings,''
JHEP {\bf 0209}, 061 (2002)
[arXiv:hep-th/0209034].

K.~Hashimoto and S.~Terashima,
``Brane decay and death of open strings,''
JHEP {\bf 0406}, 048 (2004)
[arXiv:hep-th/0404237].

\bibitem{Sen:TimeEvolution}

A.~Sen,
``Time evolution in open string theory,''
JHEP {\bf 0210}, 003 (2002)
[arXiv:hep-th/0207105].

\bibitem{Cline:DbraneCondensation}

J.~M.~Cline and H.~Firouzjahi,
``Real-time D-brane condensation,''
Phys.\ Lett.\ B {\bf 564}, 255 (2003)
[arXiv:hep-th/0301101].

\bibitem{Reheating2}

N.~Barnaby and J.~M.~Cline,
``Creating the universe from brane-antibrane annihilation,''
Phys.\ Rev.\ D {\bf 70}, 023506 (2004)
[arXiv:hep-th/0403223].

\bibitem{ClosedStrings}

A.~Sen,
``Open and closed strings from unstable D-branes,''
Phys.\ Rev.\ D {\bf 68}, 106003 (2003)
[arXiv:hep-th/0305011].

A.~Sen,
``Open-closed duality at tree level,''
Phys.\ Rev.\ Lett.\  {\bf 91}, 181601 (2003)
[arXiv:hep-th/0306137].

A.~Sen,
``Open-closed duality: Lessons from matrix model,''
Mod.\ Phys.\ Lett.\ A {\bf 19}, 841 (2004)
[arXiv:hep-th/0308068].

H.~U.~Yee and P.~Yi,
``Open / closed duality, unstable D-branes, and coarse-grained closed
strings,''
Nucl.\ Phys.\ B {\bf 686}, 31 (2004)
[arXiv:hep-th/0402027].

\bibitem{TachyonicPreheating1}

G.~N.~Felder, J.~Garcia-Bellido, P.~B.~Greene, L.~Kofman, A.~D.~Linde and I.~Tkachev,
``Dynamics of symmetry breaking and tachyonic preheating,''
Phys.\ Rev.\ Lett.\  {\bf 87}, 011601 (2001)
[arXiv:hep-ph/0012142].

G.~N.~Felder, L.~Kofman and A.~D.~Linde,
``Tachyonic instability and dynamics of spontaneous symmetry breaking,''
Phys.\ Rev.\ D {\bf 64}, 123517 (2001)
[arXiv:hep-th/0106179].

\bibitem{TachyonicPreheating2}

E.~J.~Copeland, S.~Pascoli and A.~Rajantie,
``Dynamics of tachyonic preheating after hybrid inflation,''
Phys.\ Rev.\ D {\bf 65}, 103517 (2002)
[arXiv:hep-ph/0202031].

\bibitem{DBIAction}

M.~R.~Garousi,
``Tachyon couplings on nonBPS D-branes and Dirac-Born-Infeld action,''
Nucl.\ Phys.\ B {\bf 584}, 284 (2000)
[arXiv:hep-th/0003122].

E.~A.~Bergshoeff, M.~de Roo, T.~C.~de Wit, E.~Eyras and S.~Panda,
``T-duality and actions for nonBPS D-branes,''
JHEP {\bf 0005}, 009 (2000)
[arXiv:hep-th/0003221].

J.~Kluson,
``Proposal for non-BPS D-brane action,''
Phys.\ Rev.\ D {\bf 62}, 126003 (2000)
[arXiv:hep-th/0004106].

A.~Sen,
``Tachyon matter,''
JHEP {\bf 0207}, 065 (2002)
[arXiv:hep-th/0203265].

\bibitem{Sen:KinkAndVortex}

A.~Sen,
``Dirac-Born-Infeld action on the tachyon kink and vortex,''
Phys.\ Rev.\ D {\bf 68}, 066008 (2003)
[arXiv:hep-th/0303057].

\bibitem{DBIderive}

D.~Kutasov and V.~Niarchos,
``Tachyon effective actions in open string theory,''
Nucl.\ Phys.\ B {\bf 666}, 56 (2003)
[arXiv:hep-th/0304045].

M.~Smedback,
``On effective actions for the bosonic tachyon,''
JHEP {\bf 0311}, 067 (2003)
[arXiv:hep-th/0310138].

\bibitem{DBIproperties}

A.~Sen,
``Rolling tachyon,''
JHEP {\bf 0204}, 048 (2002)
[arXiv:hep-th/0203211].

N.~Lambert, H.~Liu and J.~Maldacena,
``Closed strings from decaying D-branes,''
arXiv:hep-th/0303139.

F.~Larsen, A.~Naqvi and S.~Terashima,
``Rolling tachyons and decaying branes,''
JHEP {\bf 0302}, 039 (2003)
[arXiv:hep-th/0212248].

\bibitem{Ishida:RollingDownToDbrane}

A.~Ishida and S.~Uehara,
``Rolling down to D-brane and tachyon matter,''
JHEP {\bf 0302}, 050 (2003)
[arXiv:hep-th/0301179].

\bibitem{VacuumDBI}

G.~W.~Gibbons, K.~Hori and P.~Yi,
``String fluid from unstable D-branes,''
Nucl.\ Phys.\ B {\bf 596}, 136 (2001)
[arXiv:hep-th/0009061].

G.~Gibbons, K.~Hashimoto and P.~Yi,
``Tachyon condensates, Carrollian contraction of Lorentz group, and fundamental
strings,''
JHEP {\bf 0209}, 061 (2002)
[arXiv:hep-th/0209034].

O.~K.~Kwon and P.~Yi,
``String fluid, tachyon matter, and domain walls,''
JHEP {\bf 0309}, 003 (2003)
[arXiv:hep-th/0305229].

\bibitem{Caustics1}

G.~N.~Felder, L.~Kofman and A.~Starobinsky,
``Caustics in tachyon matter and other Born-Infeld scalars,''
JHEP {\bf 0209}, 026 (2002)
[arXiv:hep-th/0208019].

\bibitem{Caustics2}

G.~N.~Felder and L.~Kofman,
``Inhomogeneous fragmentation of the rolling tachyon,''
Phys.\ Rev.\ D {\bf 70}, 046004 (2004)
[arXiv:hep-th/0403073].

\bibitem{Caustics3}

N.~Barnaby,
``Caustic formation in tachyon effective field theories,''
JHEP {\bf 0407}, 025 (2004)
[arXiv:hep-th/0406120].

\bibitem{Tye:ImprovedAction}

N.~T.~Jones and S.~H.~H.~Tye,
``An improved brane anti-brane action from boundary superstring field theory
and multi-vortex solutions,''
JHEP {\bf 0301}, 012 (2003)
[arXiv:hep-th/0211180].

\bibitem{BSFT}

D.~Kutasov, M.~Marino and G.~W.~Moore,
``Remarks on tachyon condensation in superstring field theory,''
arXiv:hep-th/0010108.

P.~Kraus and F.~Larsen,
``Boundary string field theory of the DD-bar system,''
Phys.\ Rev.\ D {\bf 63}, 106004 (2001)
[arXiv:hep-th/0012198].

T.~Takayanagi, S.~Terashima and T.~Uesugi,
``Brane-antibrane action from boundary string field theory,''
JHEP {\bf 0103}, 019 (2001)
[arXiv:hep-th/0012210].

\bibitem{Garousi:VortexAction}

M.~R.~Garousi,
``D-brane anti-D-brane effective action and brane interaction in open string
channel,''
arXiv:hep-th/0411222.

\bibitem{StaticKinks}

P.~Brax, J.~Mourad and D.~A.~Steer,
``Tachyon kinks on non BPS D-branes,''
Phys.\ Lett.\ B {\bf 575}, 115 (2003)
[arXiv:hep-th/0304197].

E.~J.~Copeland, P.~M.~Saffin and D.~A.~Steer,
``Singular tachyon kinks from regular profiles,''
Phys.\ Rev.\ D {\bf 68}, 065013 (2003)
[arXiv:hep-th/0306294].

P.~Brax, J.~Mourad and D.~A.~Steer,
``On tachyon kinks from the DBI action,''
arXiv:hep-th/0310079.

\bibitem{DescentRelations}

J.~A.~Minahan and B.~Zwiebach,
``Effective tachyon dynamics in superstring theory,''
JHEP {\bf 0103}, 038 (2001)
[arXiv:hep-th/0009246].

K.~Hashimoto and S.~Nagaoka,
``Realization of brane descent relations in effective theories,''
Phys.\ Rev.\ D {\bf 66}, 026001 (2002)
[arXiv:hep-th/0202079].

\bibitem{Shellard}
A.~Avgoustidis and E.~P.~S.~Shellard,
``Cosmic string evolution in higher dimensions,''
arXiv:hep-ph/0410349.

\bibitem{Sakel}
M.~Sakellariadou,
``A note on the evolution of cosmic string / superstring networks,''
arXiv:hep-th/0410234.

\bibitem{DamourVilenkin1}
T.~Damour and A.~Vilenkin,
``Gravitational radiation from cosmic (super)strings: Bursts, stochastic
background, and observational windows,''
arXiv:hep-th/0410222;

\bibitem{friction}

C.~J.~A.~Martins and E.~P.~S.~Shellard,
``Quantitative String Evolution,''
Phys.\ Rev.\ D {\bf 54}, 2535 (1996)
[arXiv:hep-ph/9602271].

C.~J.~A.~Martins and E.~P.~S.~Shellard,
``Extending the velocity-dependent one-scale string evolution model,''
Phys.\ Rev.\ D {\bf 65}, 043514 (2002)
[arXiv:hep-ph/0003298].

\bibitem{DamourVilenkin2}
``Gravitational wave bursts from cusps and kinks on cosmic strings,''
Phys.\ Rev.\ D {\bf 64}, 064008 (2001)
[arXiv:gr-qc/0104026];
T.~Damour and A.~Vilenkin,
``Gravitational wave bursts from cosmic strings,''
Phys.\ Rev.\ Lett.\  {\bf 85}, 3761 (2000)
[arXiv:gr-qc/0004075].

\bibitem{MajumDavis}
M.~Majumdar and A.~C.~Davis,
``D-brane anti-brane annihilation in an expanding universe,''
JHEP {\bf 0312}, 012 (2003)
[arXiv:hep-th/0304153].

\bibitem{GarHind}
T.~Garagounis and M.~Hindmarsh,
``Scaling in numerical simulations of domain walls,''
Phys.\ Rev.\ D {\bf 68}, 103506 (2003)
[arXiv:hep-ph/0212359].

\bibitem{OMA}
J.~C.~R.~Oliveira, C.~J.~A.~Martins and P.~P.~Avelino,
``The cosmological evolution of domain wall networks,''
arXiv:hep-ph/0410356.

\bibitem{vilsmith}
A.~G.~Smith and A.~Vilenkin,
``Numerical Simulation Of Cosmic String Evolution In Flat Space-Time,''
Phys.\ Rev.\ D {\bf 36}, 990 (1987).

M.~Sakellariadou and A.~Vilenkin,
``Cosmic-String Evolution In Flat Space-Time,''
Phys.\ Rev.\ D {\bf 42}, 349 (1990).

\bibitem{SF}

D.~Stojkovic and K.~Freese,
``A black hole solution to the cosmological monopole problem,''
arXiv:hep-ph/0403248.

\bibitem{frustrated}
C.~J.~A.~Martins,
``Scaling laws for non-intercommuting cosmic string networks,''
Phys.\ Rev.\ D {\bf 70}, 107302 (2004)
[arXiv:hep-ph/0410326].

\bibitem{polchinski-review}
J.~Polchinski,
``Cosmic superstrings revisited,''
arXiv:hep-th/0410082.

\bibitem{Brandenberger2}

R.~H.~Brandenberger,
``Quantum Field Theory Methods And Inflationary Universe Models,''
Rev.\ Mod.\ Phys.\  {\bf 57}, 1 (1985).

\bibitem{JST2}

N.~Jones, H.~Stoica and S.~H.~H.~Tye,
``Brane interaction as the origin of inflation,''
JHEP {\bf 0207}, 051 (2002)
[arXiv:hep-th/0203163].

\bibitem{Kawasaki}

S.~Kasuya and M.~Kawasaki,
``Can topological defects be formed during preheating?,''
Phys.\ Rev.\ D {\bf 56}, 7597 (1997)
[arXiv:hep-ph/9703354].


\end{thebibliography}
\end{document}